\documentclass[preprint,12pt]{elsarticle}
\usepackage{graphicx}
\usepackage{amsfonts}
\usepackage{amsmath}
\usepackage{url}

\def\erf{{\rm erf}}
\def\var{{\rm var}}

\def\PNL{{\mathcal P}}   
\def\dPNL{\delta\PNL}    
\def\EE{{\bf E}}
\def\A{\EE}      
\def\C{{\bf C}}
\def\I{{\bf I}}
\def\M{{\bf M}}
\def\O{{\bf O}}
\def\U{{\bf U}}

\def\N{{\mathcal N}}
\def\ve{\varepsilon}
\def\tr{{\rm tr}}

\def\r{{\bf r}}
\def\s{{\bf s}}
\def\a{{\bf a}}
\def\t{{\bar{t}}}
\def\T{{\dagger}}
\def\TO{{\mathcal T}}
\def\q{\tilde{q}}
\def\p{\tilde{p}}

\journal{Physica A}

\begin{document}

\begin{frontmatter}
\title{Following a Trend with an Exponential Moving Average: Analytical Results for a Gaussian Model}

\author{Denis~S.~Grebenkov}
\address{
Laboratoire de Physique de la Mati\`{e}re Condens\'{e}e, \\ CNRS --
Ecole Polytechnique, 91128 Palaiseau, France}
\ead{denis.grebenkov@polytechnique.edu}

\author{Jeremy~Serror}
\address{
John Locke Investment, 38 Avenue Franklin Roosevelt, 77210 Fontainebleau-Avon, France}
  \ead{jeremy.serror@jl-investments.com}

\date{\today}

\begin{abstract}
We investigate how price variations of a stock are transformed into
profits and losses (P\&Ls) of a trend following strategy.  In the
frame of a Gaussian model, we derive the probability distribution of
P\&Ls and analyze its moments (mean, variance, skewness and kurtosis)
and asymptotic behavior (quantiles).  We show that the asymmetry of
the distribution (with often small losses and less frequent but
significant profits) is reminiscent to trend following strategies and
less dependent on peculiarities of price variations.  At short times,
trend following strategies admit larger losses than one may anticipate
from standard Gaussian estimates, while smaller losses are ensured at
longer times.  Simple explicit formulas characterizing the
distribution of P\&Ls illustrate the basic mechanisms of momentum
trading, while general matrix representations can be applied to
arbitrary Gaussian models.  We also compute explicitly annualized risk
adjusted P\&L and strategy turnover to account for transaction
costs.  We deduce the trend following optimal timescale and its
dependence on both auto-correlation level and transaction costs.
Theoretical results are illustrated on the Dow Jones index.
\end{abstract}




\end{frontmatter}

\section{Introduction}

\begin{figure}
\begin{center}
\includegraphics[width=67.5mm]{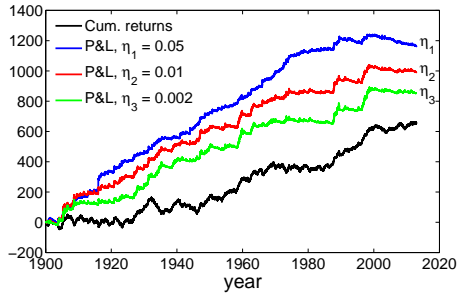}
\end{center}
\caption{
Cumulative standardized logarithmic returns (normalized by realized
volatility) of the Dow Jones index, from 1900 to 2012 (black curve),
and cumulative P\&Ls of trend following strategies (defined in
Sec. \ref{sec:market}) with timescales $\eta_1 = 0.05$ (blue), $\eta_2
= 0.01$ (red), and $\eta_3 = 0.002$ (green), applied to this index.}
\label{fig:DJ_price}
\end{figure}

Systematic trading has grown as an industry in finance, allowing to
take rapid trading decisions for multiple stocks
\cite{Covel,Clenow,Chan96,Chan99,Jegadeesh01,Moskowitz12,Asness13}.
A strategy relies on price time series in the past in order to
forecast price variations in near future and update accordingly its
positions.  Although the market complexity, variability and
stochasticity damn such forecasting to fail in nearly half cases, even
a tiny excess of successful forecasts is enhanced by a very large
number of trades into statistically relevant profits.  Many trading
strategies attempt to detect an eventual trend in price series, i.e.,
a sequence of positively auto-correlated price variations which may be
caused, e.g., by a news release or common activity of multiple
traders.  From a practical point of view, a strategy transforms the
known past information into a signal for buying or selling a number of
shares.  From a mathematical point of view, systematic trading can be
seen as a transformation of price time series into profit-and-loss
(P\&L) time series of the strategy, as illustrated on
Fig. \ref{fig:DJ_price}.  For instance, the passive (long) strategy of
buying and holding a stock corresponds to the identity transformation.
The choice for the optimal strategy depends on the imposed risk-reward
criteria.

\begin{figure}
\begin{center}
\includegraphics[width=67.5mm]{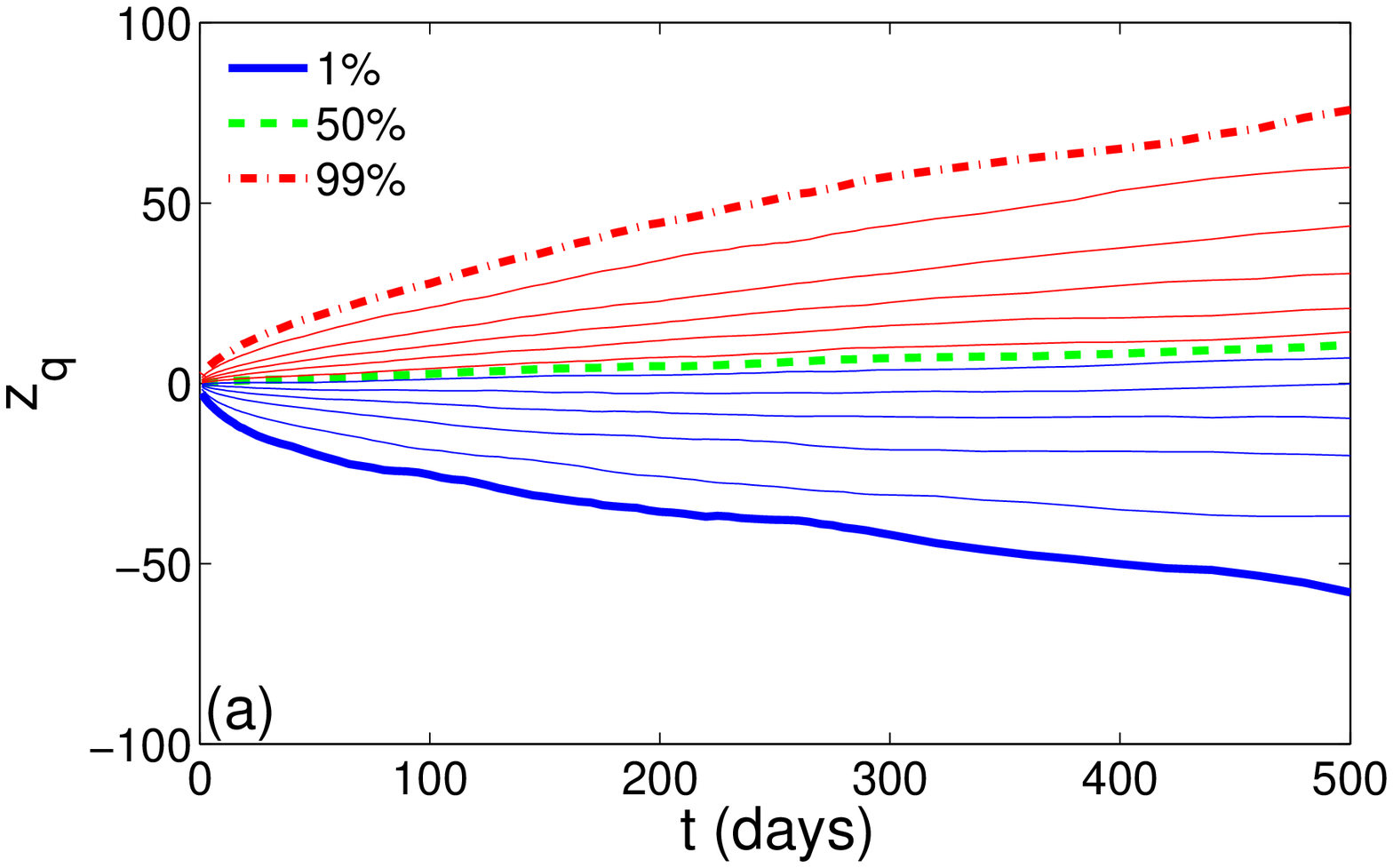} 
\includegraphics[width=67.5mm]{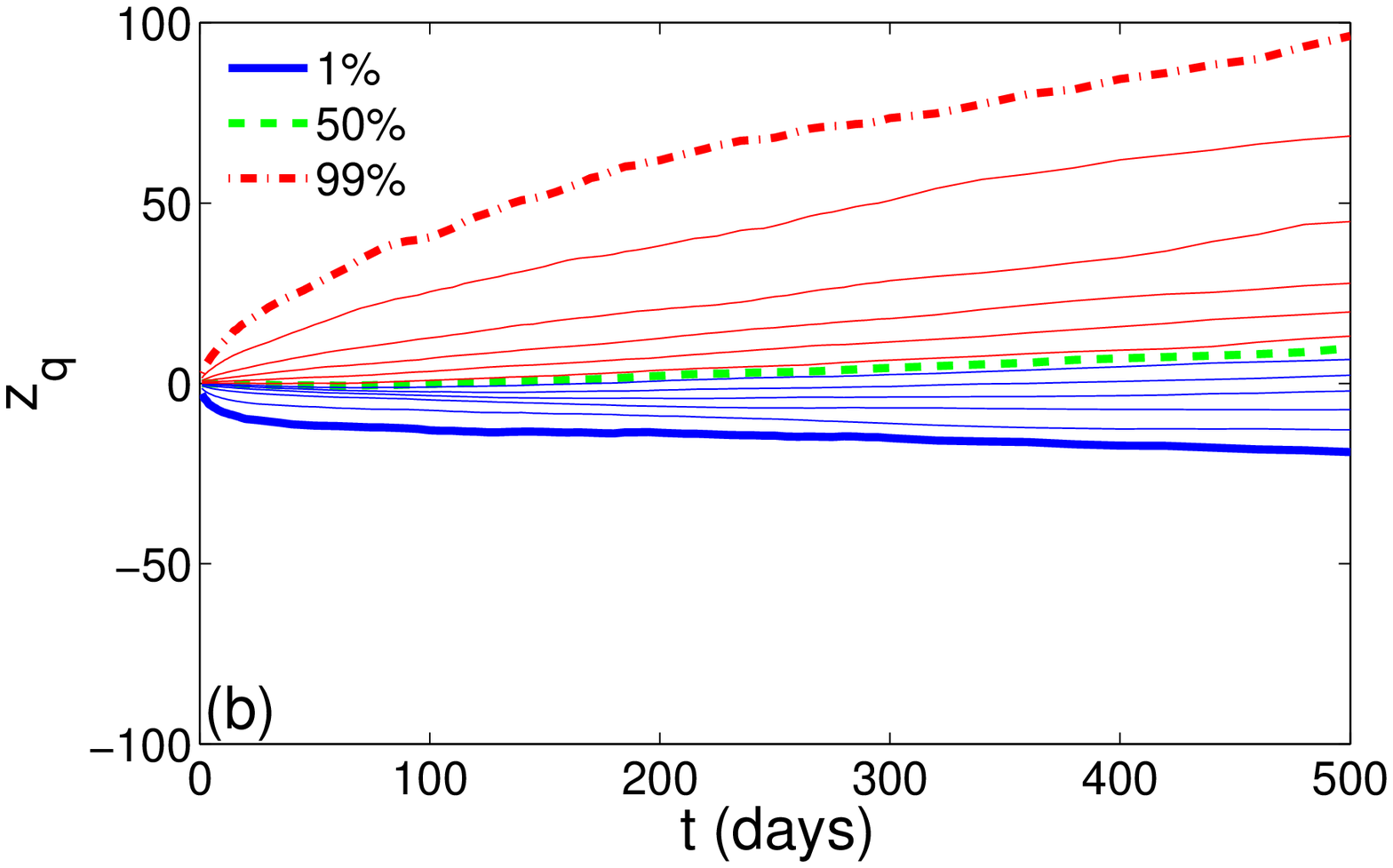}
\end{center}
\caption{
Comparison between quantiles of buy-and-hold strategy and trend
following strategy. {\bf (a)} Quantiles of cumulative standardized
logarithmic returns (normalized by realized volatility) of the Dow
Jones index (1900-2012) as functions of lag time $t$ (in days).  Lines
from the bottom to the top show quantiles with $1\%$, $5\%$, $15\%$,
$25\%$, $35\%$, $45\%$ (blue color), $50\%$ (green color), $55\%$,
$65\%$, $75\%$, $85\%$, $95\%$, and $99\%$ (red color), respectively.
Thick lines highlight $1\%$, $50\%$ and $99\%$ quantiles.  {\bf (b)}
Quantiles of cumulative P\&Ls of a trend following strategy with the
timescale $\eta = 0.01$, applied to the Dow Jones index (same
notations).}
\label{fig:DJ_quantiles}
\end{figure}

In this paper, we study the transformation of price variations into
P\&Ls of a trend following strategy based on an exponential moving
average (EMA).  This archetypical strategy turns out to be the basis
for many systematic trading platforms
\cite{Covel,Clenow,Chan96,Chan99,Jegadeesh01,Moskowitz12,Asness13},
while other methods such as the detrending moving average analysis or
higher-order moving averages can also be employed
\cite{Vandewalle98,Vandewalle99,Carbone04,Arianos11}.  A trend
following strategy is known to skew the probability distribution of
P\&Ls \cite{Potters05,Martin12}, as we illustrate on
Fig. \ref{fig:DJ_quantiles}.  This figure shows how empirically
computed quantiles of price variations%
\footnote{
Here, by price variations we mean cumulative standardized logarithmic
returns (normalized by realized volatility), to get closer to the
Gaussian hypothesis \cite{Andersen00}.}
are transformed into quantiles of P\&Ls for the Dow Jones index
(1900-2012).  Even for such a long sample with 30733 daily returns,
accurate estimation of quantiles remains problematic.  Moreover, the
basic mechanisms of this transformation remain poorly understood.  For
these reasons, we will study a simple model in which standardized
logarithmic returns are Gaussian random variables \cite{Andersen00}
whose auto-correlations reflect random trends.  Even though heavy
tailed asymptotic distribution of returns and some other stylized
facts are ignored
\cite{Bouchaud,Mantegna,Mantegna95,Gabaix03,Bouchaud01,Sornette03,Bouchaud04,Stella08,Bouchaud01b,Valeyre13},
the Gaussian hypothesis will allow us to derive analytical results
that can be later confronted to empirical market data.  We will
compute the probability distribution of P\&Ls of a trend following
strategy in order to understand how the Gaussian distribution of price
variations is transformed by systematic trading.  The respective roles
of the market (positive auto-correlations) and of the strategy itself,
onto profits and losses, will therefore be disentangled.

The paper is organized as follows.  In Sec. \ref{sec:market}, we
introduce matrix notations, a market model and a trend following
strategy.  Main results about the probability distribution and moments
of P\&Ls are presented in Sec. \ref{sec:PNL}.  Discussion, conclusion
and perspectives are summarized in Sec. \ref{sec:discussion}.

\section{Market model and trading strategy}
\label{sec:market}

\subsection{Exponential moving average}

The exponential moving average (EMA) is broadly employed in signal
processing and data analysis \cite{Box,Holt57,Winters60,Brown}.  The
EMA can be defined as a linear transformation of a time series
$\{x_t\}$ to a smoother time series $\{\tilde{x}_t\}$ according to
\begin{equation}
\label{eq:EMA}
\tilde{x}_t = \lambda \sum\limits_{k=0}^{\infty} (1-\lambda)^k x_{t-k} ,
\end{equation}
where $0 < \lambda \leq 1$ is the (inverse of) timescale.  When
$\lambda = 1$, the EMA is the identity transformation: $\tilde{x}_t =
x_t$; in contrast, many terms $x_{t-k}$ effectively contribute to
$\tilde{x}_t$ when $\lambda\ll 1$.  The EMA is often preferred to
simple moving average over a window of fixed length because it yields
smoother results.  In practice, it can be computed in real time
according to a recurrent formula:
\begin{equation*}
\tilde{x}_t = (1-\lambda) \tilde{x}_{t-1} + \lambda x_t .
\end{equation*}
When a time series starts from $t = 1$, the non-existing elements
$x_0$, $x_{-1}$, $x_{-2}$, ... are set to $0$.  This is equivalent to
setting the upper limit in Eq. (\ref{eq:EMA}) to $t-1$.  In the
analysis of a finite sample of length $T$, the EMA can be written in a
matrix form as
\begin{equation*}
\left(\begin{array}{c} \tilde{x}_1 \\ ... \\ \tilde{x}_T \\ \end{array}\right) = \lambda \EE_{1-\lambda} 
\left(\begin{array}{c} x_1 \\ ... \\ x_T \\ \end{array}\right) ,
\end{equation*}
where $\EE_q$ is the matrix of size $T\times T$, whose elements are
\begin{equation}
(\EE_q)_{jk} = \begin{cases} q^{j-k-1}  \quad (j>k), \cr
 0 \hskip 13mm  (j\leq k). \end{cases}
\end{equation}

\subsection{Gaussian market models}

The first Gaussian market model was introduced by Bachelier in 1900
and since that time, numerous models have been developed.  For
instance, the class of ARMA (Auto-Regressive Moving Average) models
and their extensions were thoroughly employed in finance \cite{Box}.
During decades, these models were getting more and more elaborate in
order to account for various empirical features of markets.  Our
purpose is the opposite: we aim at understanding the basic mechanisms
of trend following strategies, and we expect that qualitatively, these
mechanisms weakly depend on market peculiarities.  In turn, the
quantitative behavior of P\&Ls may of course be sensitive to
particular features.  For this reason, we choose a simple model
exhibiting random trends, in order to be able to derive analytical
results.  At the same time, general matrix formulas used in this paper
(see the beginning of Sec. \ref{sec:PNL}) can be applied to arbitrary
Gaussian market model.  In this light, our methodology can be used for
studying more elaborate models, though results will be less explicit.

\begin{figure}
\begin{center}
\includegraphics[width=67.5mm]{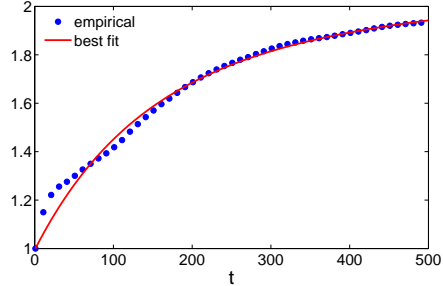}
\end{center}
\caption{
Variogram of standardized logarithmic daily returns (normalized by
realized volatility) of the Dow Jones index (full circles) and its
best fit to Eq. (\ref{eq:Vst_exo}) (red line) with $\lambda = 0.011$
and $\beta_0 = 0.08$. }
\label{fig:DJ_variogram}
\end{figure}

In this paper, we consider a simple model of daily price variations,
or returns,%
\footnote{
Throughout this paper, daily price variations will be called
``returns'' for the sake of simplicity.  Rigorously speaking, we
consider additive standardized logarithmic returns normalized by
realized volatility.  Such a resizing, which is a common practice on
futures markets \cite{Martin12}, allows one to reduce, to some extent,
the impact of changes in volatility and its correlations
\cite{Bouchaud01b,Valeyre13}, and to get closer to the Gaussian
hypothesis of returns \cite{Andersen00}.}
$r_t$, in which random trends are induced by a discrete
Ornstein-Uhlenbeck process, while short-time fluctuations are modeled
by iid Gaussian variables $\ve_k\in \N(0,1)$ with zero mean and unit
variance:
\begin{equation}
\label{eq:rt_exo}
r_t = \ve_t + \beta \sum\limits_{k=1}^{t-1} (1-\lambda)^{t-1-k} \xi_k ,
\end{equation}
where $\lambda$ and $\beta$ are two parameters of the market model
describing the characteristic timescale and the strength of the trend
contribution, and $\xi_k\in \N(0,1)$ are iid Gaussian variables
(independent of $\ve_k$).  This is a model of stochastic trends which
are induced by a persistent process generated by exogeneous random
variables $\xi_k$ which are independent from the short-time
fluctuations $\ve_k$.  In a matrix form, one writes
\begin{equation}
\r = \ve + \beta \EE_{1-\lambda} \xi ,
\end{equation}
where $\r = (r_1,...,r_T)^\T$ is the vector of returns (the
superscript $\T$ denoting the transpose), and $\ve$ and $\xi$ are two
vectors of iid Gaussian variables.  As a consequence, $\r$ is a
Gaussian vector with zero mean, $\langle r_j\rangle = 0$, and the
covariance matrix $\C_{j,k} = \langle r_j r_k\rangle$, for which
\begin{equation}
\label{eq:C1}
\C = \I + \beta^2 \EE_{1-\lambda} \EE_{1-\lambda}^\T ,
\end{equation}
where $\I$ stands for the identity matrix.  The elements of this
matrix are
\begin{equation}
\C_{j,k} = \delta_{j,k} + \frac{\beta^2}{\lambda(2-\lambda)} \biggl[(1-\lambda)^{|j-k|} - (1-\lambda)^{j+k-2}\biggr] .
\end{equation}
The second term is the covariance of a discrete Ornstein-Uhlenbeck
process.  The diagonal elements $\C_{t,t}$ approach the constant
$\sigma_\infty^2 = 1 + \frac{\beta^2}{\lambda(2-\lambda)}$ as
$t\to\infty$, i.e., auto-correlations increase the variance of
returns.  It is convenient to make the limiting variance
$\sigma_\infty^2$ independent of the timescale $\lambda$ by rescaling
the parameter $\beta$ as
\begin{equation}
\label{eq:beta}
\beta = \beta_0\sqrt{\lambda(2-\lambda)} ,
\end{equation}
so that $\sigma_\infty^2 = 1 + \beta_0^2$, independently of $\lambda$.
In other words, the new parameter $\beta_0^2$ is the asymptotic excess
variance of returns due to their auto-correlations.  This parameter
can be calibrated from empirical price series.  For this purpose, we
consider the variogram of returns over the lag time $t$
\begin{equation}
V_{t,t_0} \equiv \frac{\var\{r_{t_0+1} + ... + r_{t_0+t}\}}{\var\{r_{t_0+1}\} + ... + \var\{r_{t_0+t}\}} ,
\end{equation}
where an initiation period of duration $t_0$ can be ignored to
approach the stationary regime.  The variogram would be equal to $1$
for iid random variables, while its deviations from $1$ characterize
auto-correlations between variables.  Expressing the variogram through
the covariance matrix in Eq. (\ref{eq:C1}), one gets in the stationary
limit $t_0\to\infty$:
\begin{equation*}
\lim\limits_{t_0\to\infty} \var\{r_{t_0+1} + ... + r_{t_0+t} \} = t \biggl[1 + \beta_0^2 \frac{2-\lambda}{\lambda}\biggr] 
- \frac{2(1-\lambda)\beta_0^2}{\lambda^2} \bigl(1-(1-\lambda)^t\bigr)
\end{equation*}
and
\begin{equation}
\label{eq:Vst_exo}
V_{t,\infty} = 1 + \frac{2(1-\lambda)\beta_0^2}{\lambda(1 + \beta_0^2)}
\biggl(1 - \frac{1 - (1-\lambda)^{t}}{\lambda t}\biggr) .
\end{equation}
Figure \ref{fig:DJ_variogram} shows the empirical variogram of returns
obtained from the Dow Jones index (1900-2012), and its fit according
to Eq. (\ref{eq:Vst_exo}).  Although the model fails to reproduce a
steep growth of auto-corrections at short times, it captures correctly
the behavior of the variogram at longer times and allows us to get
realistic values for the parameters $\lambda$ and $\beta_0$ of the
model: $\lambda = 0.011$ and $\beta_0 = 0.08$.  At the same time,
these values are market dependent and, in general, difficult to
calibrate.  In what follows, the representative values $\lambda =
0.01$ and $\beta_0 = 0.1$ will be used for illustrative purposes.

For comparison, we also consider an auto-regressive model with
exponential weights,
\begin{equation}
\label{eq:rt_endo}
r_t = \ve_t + \beta\sum\limits_{k=1}^{t-1} (1-\lambda)^{t-1-k} r_k ,
\end{equation}
where $\ve_t\in\N(0,1)$ are iid Gaussian variables.  This is a model
of autoregressive trends which are induced through auto-correlations
with earlier returns.  Writing Eq. (\ref{eq:rt_endo}) in a matrix
form, $\r = \ve + \beta \EE_{1-\lambda} \r$ and inverting this
relation yields
\begin{equation}
\r = (\I - \beta \EE_{1-\lambda})^{-1} \ve = (\I + \beta \EE_{1-\lambda+\beta}) \ve ,
\end{equation}
where the explicit matrix inversion was possible due to the specific
triangular structure of the matrix $\EE_q$.  As a consequence, $\r$ is
a Gaussian vector with zero mean and the covariance matrix
\begin{equation}
\label{eq:C2}
\C = (\I + \beta \EE_{\q}) (\I + \beta \EE_{\q})^\T = \I + \beta(\EE_{\q} + \EE_{\q}^\T) + \beta^2 \EE_{\q} \EE_{\q}^\T,
\end{equation}
where $\q = 1 - \lambda + \beta$.  Comparing this relation to
Eq. (\ref{eq:C1}), one notes the effective timescale $\lambda-\beta$
(instead of $\lambda$), and an additional term $\beta(\EE_{\q} +
\EE_{\q}^\T)$.  Although both models exhibit many similar features,
they are not identical due to the presence of this term.  For the sake
of simplicity, we focus on the stochastic trend model (defined by
Eq. (\ref{eq:rt_exo})), while similar results for the auto-regressive
trend model are derived and discussed in \cite{SM}.  Qualitative
conclusions of the paper do not depend on this choice.

It is worth noting that we focus on trends that are induced by
auto-correlations, while mean returns are zero.  While an extension to
the case $\langle r_t\rangle \ne 0$ is relatively straightforward, the
studied situation with $\langle r_t\rangle = 0$ allows us to easier
illustrate the role of trend following strategy because the passive
holding strategy is profitless in this case.

\subsection{Trading strategy}
\label{sec:strategy}

The trading strategy relies on an EMA of returns in order to detect
eventual trends in price time series
\cite{Box,Holt57,Winters60,Brown}.  We consider the signal $s_t$,
which is proportional to the EMA of returns:
\begin{equation}
\label{eq:St_prop}
s_t = \gamma \sum\limits_{k=1}^{t-1} (1-\eta)^{t-1-k} r_k ,
\end{equation}
where $\eta$ and $\gamma$ are two parameters of the strategy (in what
follows, we will relate $\gamma$ to $\eta$, the latter remaining the
only parameter of the strategy).  It is crucial that the signal at
time $t$ is determined by {\it earlier} returns $r_{t-1}$, $r_{t-2}$,
... and does not rely on unavailable information on the present return
$r_t$.  In a matrix form, Eq. (\ref{eq:St_prop}) reads as
\begin{equation}
\s = \gamma \EE_{1-\eta} \r  .
\end{equation}

The cumulative P\&L of a trend following strategy after $t$ steps is
defined as
\begin{equation}
\label{eq:PNL}
\PNL_{t,t_0} \equiv \sum\limits_{k=t_0+1}^{t+t_0} r_k s_k = (\r^\T \O^{(t,t_0)} \s) 
= \gamma (\r^\T \O^{(t,t_0)} \A_{1-\eta} \r) = \frac12 (\r^\T \M^{(t,t_0)}_{1-\eta} \r) ,
\end{equation}
where $t_0$ is the duration of an initiation period,
\begin{equation}
\label{eq:Mt}
\M_{1-\eta}^{(t,t_0)} \equiv \gamma\bigl[\O^{(t,t_0)}\A_{1-\eta} + \A_{1-\eta}^\T \O^{(t,t_0)}\bigr]
\end{equation}
is a symmetric matrix, and $\O^{(t,t_0)}$ is the matrix which has $1$
in the diagonal positions between $t_0+1$ and $t_0+t$, and $0$
elsewhere.  The cumulative P\&L in Eq. (\ref{eq:PNL}) is written as a
quadratic form of the Gaussian vector $\r$.  Similarly, an incremental
P\&L reads as
\begin{equation}
\label{eq:dPNL}
\dPNL_\t \equiv \dPNL_{t,t_0} \equiv \PNL_{t,t_0} - \PNL_{t-1,t_0} = r_{t+t_0} s_{t+t_0} 
= \frac12 (\r^\T \M_{1-\eta}^{(1,\t-1)} \r),
\end{equation}
where $\t = t + t_0$ is a shortcut notation for $t+t_0$.


\section{Profit-and-loss of trend following strategy}
\label{sec:PNL}

The representations (\ref{eq:PNL}, \ref{eq:dPNL}) of cumulative and
incremental P\&Ls as quadratic forms of Gaussian vectors allow one to
investigate their properties.  For a discrete Gaussian process $\r$
with mean zero and covariance matrix $\C$, the quadratic form $\chi =
\frac12(\r^\T \M \r)$ defined by a symmetric matrix $\M$ is a random
variable whose moments and probability distribution are well known
\cite{Grebenkov11}.  In fact, a matrix representation of the
characteristic function of $\chi$,
\begin{equation}
\phi(k) \equiv \langle \exp(ik\chi)\rangle = \frac{1}{\sqrt{\det(\I - ik\M\C)}}, 
\end{equation}
yields the probability density $p(z)$ of $\chi$ through the inverse
Fourier transform:
\begin{equation}
\label{eq:pz}
p(z) = \int\limits_{-\infty}^\infty \frac{dk}{2\pi} e^{-ikz} \phi(k) .
\end{equation}
As described in \cite{Grebenkov11}, the determinant $\det(\I -
ik\M\C)$ can be expressed through the eigenvalues of the matrix $\M\C$
that speeds up numerical computations.  Moreover, the smallest and the
largest eigenvalues, $\mu_{-}$ and $\mu_+$, essentially determine the
asymptotic behavior of the probability density $p(z)$:
\begin{equation}
\label{eq:pz_asympt}
p(z) \propto A_\pm z^{\nu_\pm} \exp(-z/\mu_\pm)  \quad (z\to \pm\infty)
\end{equation}
(note that $\mu_- < 0$ to ensure the decay of the density as $z\to
-\infty$).

Finally, the cumulant moments $\kappa_m$ of the quadratic form $\chi$
are
\begin{equation}
\label{eq:kappa}
\kappa_m = \frac{(m-1)!}{2}~ \tr((\M \C)^m),
\end{equation}
where $\tr$ denotes the trace.  In particular, $\kappa_1$ and
$\kappa_2$ are the mean and variance of $\chi$, while higher-order
cumulant moments determine the skewness ($\kappa_3/\kappa_2^{3/2}$)
and kurtosis ($\kappa_4/\kappa_2^2$).

\subsection{Mean incremental P\&L}

We first consider the incremental P\&L, $\dPNL_\t$, for which the
matrix $\M_{1-\eta}^{(1,\t-1)}$ from Eq. (\ref{eq:Mt}) has a
particularly simple structure, with nonzero contributions only at
$\t$-th row and column.  The product of this matrix with the
covariance matrix $\C$ can be written explicitly, e.g.,
\begin{equation}
\label{eq:kappa1_dPNL}
\kappa_1 = \frac12 \tr(\M\C) = \gamma \sum\limits_{k=1}^{\t-1} p^{\t-k-1} \C_{\t,k} ,
\end{equation}
where $p = 1-\eta$.  Substituting Eq. (\ref{eq:C1}) into
Eq. (\ref{eq:kappa1_dPNL}) yields the mean incremental P\&L
\begin{equation}
\label{eq:dPNL_mean_exo}
\langle \dPNL_\t\rangle = \gamma \beta_0^2 \biggl[q \frac{1 - (pq)^{\t-1}}{1-pq} - q^{\t-1} \frac{p^{\t-1} - q^{\t-1}}{p-q}\biggr] ,
\end{equation}
where $q = 1-\lambda$.  In the special case $\eta =
\lambda$, this expression reduces to
\begin{equation}
\langle \dPNL_\t\rangle = \gamma \beta_0^2 \biggl[q \frac{1 - q^{2(\t-1)}}{1-q^2} - (\t-1) q^{2\t-3}\biggr] .
\end{equation}
In the stationary limit $t_0\to\infty$ (we recall that $\t = t+t_0$),
Eq. (\ref{eq:dPNL_mean_exo}) yields the mean stationary daily P\&L:
\begin{equation}
\label{eq:dPNL_inf_exo}
\langle \dPNL_\infty\rangle = \gamma \beta_0^2 \frac{1-\lambda}{1-(1-\eta)(1-\lambda)} .
\end{equation}

The mean cumulative P\&L, $\langle \PNL_{t,t_0}\rangle$, can be
obtained by summing contributions in Eq. (\ref{eq:dPNL_mean_exo}).  In
the stationary limit $t_0\to\infty$, the mean cumulative P\&L is
simply proportional to $t$:
\begin{equation*}
\langle \PNL_{t,\infty}\rangle = t~ \langle \dPNL_{\infty}\rangle.
\end{equation*}

\subsection{Variance of incremental P\&L}

The variance $v_\t$ of the incremental P\&L, $\dPNL_\t$, is
\begin{equation*}
v_\t \equiv 
\langle r_\t^2 s_\t^2\rangle - \langle r_\t s_\t\rangle^2  
 = \gamma^2 \sum\limits_{j,k=1}^{\t-1} [\A_{1-\eta}]_{\t,j} [\A_{1-\eta}]_{\t,k} \bigl[\langle r_\t^2 r_j r_k \rangle
- \langle r_\t r_j \rangle \langle r_\t r_k \rangle\bigr] .
\end{equation*}
Since $\r$ is a Gaussian vector, the Wick's theorem allows one to
express the fourth-order correlation of Gaussian variables through the
covariance matrix:
\begin{equation*}
\langle r_\t^2 r_j r_k \rangle = \langle r_\t^2\rangle  ~ \langle r_j r_k\rangle + 2 \langle r_\t r_j\rangle \langle r_\t r_k\rangle 
= \C_{\t,\t} \C_{j,k} + 2 \C_{\t,j} \C_{\t,k} ,
\end{equation*}
from which
\begin{equation}
\label{eq:vt_general}
v_\t  = \gamma^2 \biggl[ \C_{\t,\t} (\A_{1-\eta} \C \A_{1-\eta}^\T)_{\t,\t} + [(\A_{1-\eta} \C)_{\t,\t}]^2\biggr] . 
\end{equation}
Substituting Eq. (\ref{eq:C1}) yields
\begin{equation}
\label{eq:vt}
\begin{split}
& v_\t = \gamma^2 \biggl\{ \bigl(1 + \beta_0^2 (1-q^{2\t-2})\bigr) \biggl(\frac{1-p^{2\t-2}}{1-p^2} + \frac{\beta_0^2(1-q^2)}{(1-pq)(p-q)}
\biggl[\frac{p(1 - p^{2\t-2})}{1-p^2} - \\
& \frac{q(1-q^{2\t-2})}{1-q^2} - \frac{(p^{\t-1}-q^{\t-1})^2}{p-q}\biggr] \biggr)
 + \frac{\beta_0^4 (1-q^2)^2}{(p-q)^2} \biggl[\frac{1-(pq)^{\t-1}}{1-pq} - \frac{1-q^{2\t-2}}{1-q^2}\biggr]^2 \biggr\} . \\
\end{split}
\end{equation}
In the special case $\eta = \lambda$, one gets
\begin{equation*}
\begin{split}
 v_\t = \frac{\gamma^2}{1-q^2} & \biggl\{ \bigl(1 + \beta_0^2 (1-q^{2\t-2})\bigr) \biggl( (1-q^{2\t-2}) \\
& + \frac{\beta_0^2}{1-q^2} \biggl[1+q^2 - q^{2(\t-1)} \biggl(1 + q^2\bigl[1 + (\t-1)(q^{-2}-1)\bigr]^2\biggr)\biggr] \biggr) \\
&  + \frac{\beta_0^4 q^2}{1-q^2} \biggl[1 - q^{2(\t-1)} \bigl[1 + (\t-1)(q^{-2}-1)\bigr]\biggr]^2 \biggr\} , \\
\end{split}
\end{equation*}
In the stationary limit $t_0\to\infty$, Eq. (\ref{eq:vt}) reduces to
\begin{equation}
\label{eq:v_infty}
v_\infty = \frac{\gamma^2}{1-p^2} \biggl[1 + \frac{2\beta_0^2}{1-pq} + \frac{\beta_0^4(1+q^2-2p^2q^2)}{(1-pq)^2}\biggr].
\end{equation}
Setting the parameter $\gamma$ of the strategy to
\begin{equation}
\label{eq:gamma}
\gamma^2 = 1-p^2 = \eta(2-\eta)
\end{equation}
ensures the unit variance of the incremental P\&L for the case of
independent returns (i.e., when $\beta_0=0$).  The condition allows
one to properly compare trend following and passive (long) strategies.
When $\lambda \ll 1$, $\eta \ll 1$ and $\beta_0^2 \ll 1$, one gets
$v_\infty \approx 1 + \beta_0^2 \frac{2}{\lambda + \eta}$, while the
stationary variance of returns was $\sigma_\infty^2 = 1 + \beta_0^2$.
In other words, the correction term $\beta_0^2$ is enhanced by the
large factor $\frac{2}{\lambda+\eta}$.

One can also consider the variogram of incremental P\&Ls for which we
derive in \ref{sec:variogramPNL} the exact formula in the stationary
limit $t_0\to\infty$.  Interestingly, the variogram of incremental
P\&L can be larger or smaller than the variogram of returns, depending
on the timescale $\eta$ of the strategy.  It is worth noting that the
variogram of incremental P\&Ls is equal to $1$ for the case of
independent returns.

\begin{figure}
\begin{center}
\includegraphics[width=67.5mm]{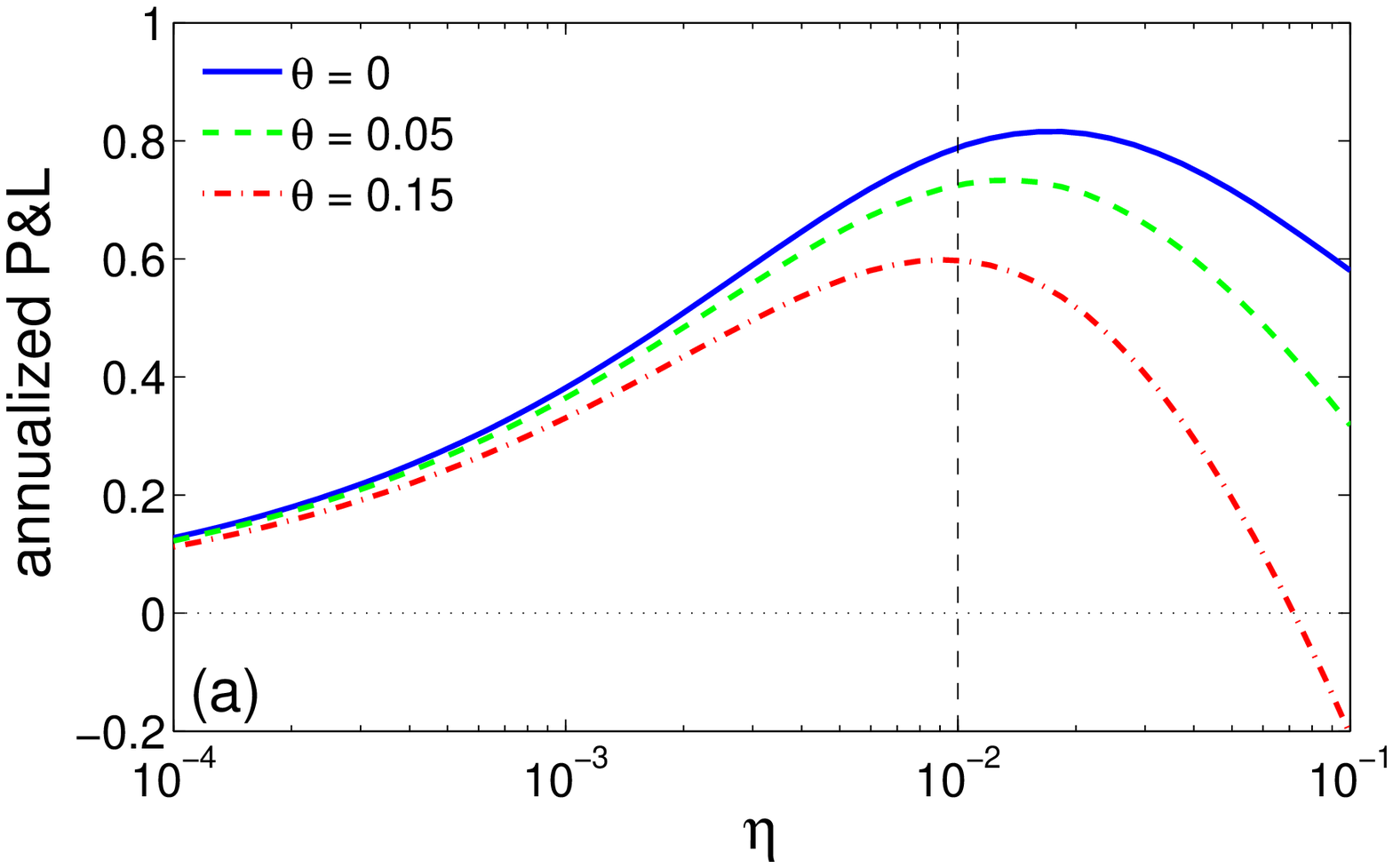}
\includegraphics[width=67.5mm]{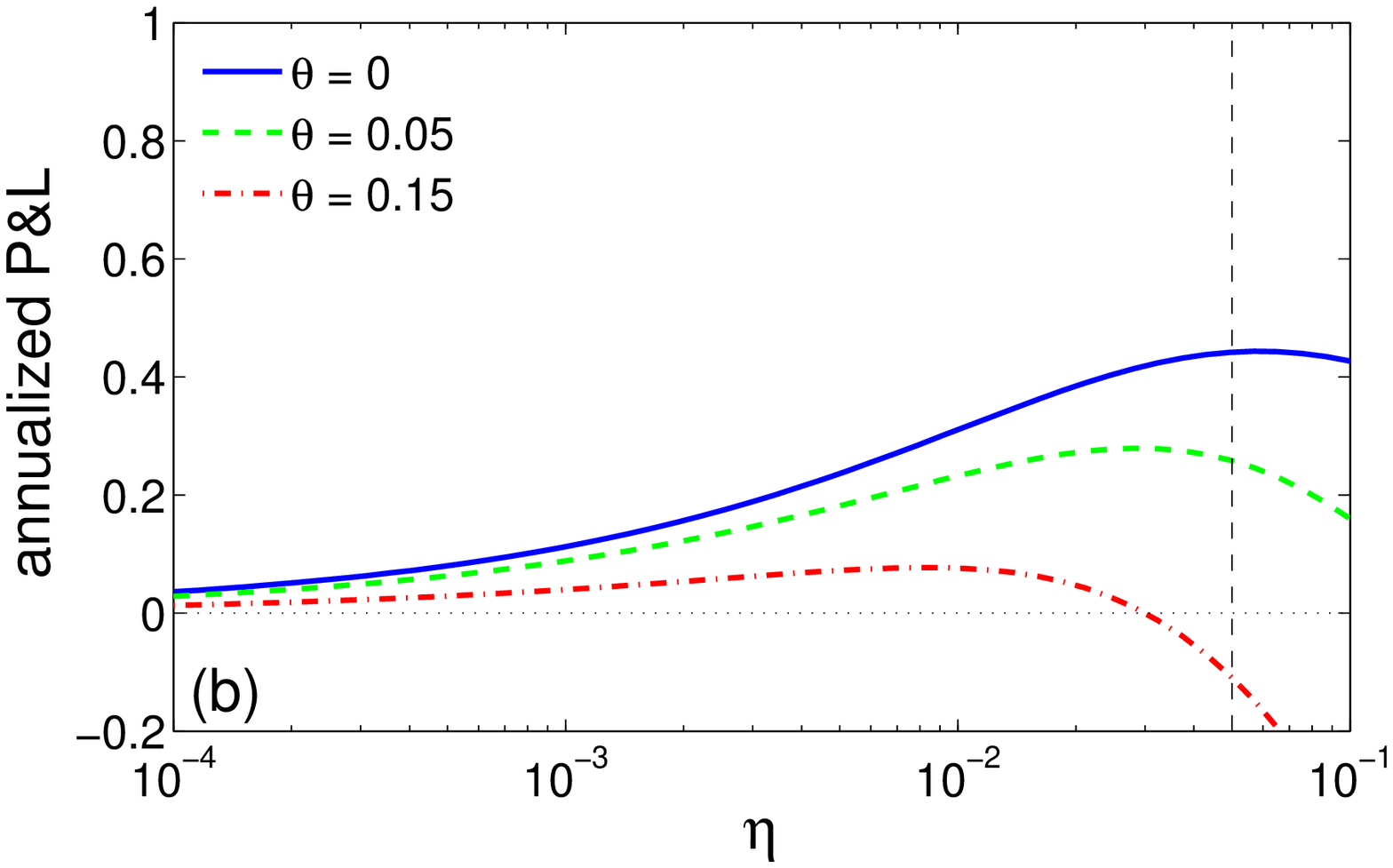}
\end{center}
\caption{
Net annualized risk adjusted P\&L, $\sqrt{255}~\frac{\langle
\dPNL_\infty\rangle - \langle \TO_\infty \rangle}{\sqrt{v_\infty}}$, as
a function of the strategy timescale $\eta$, for the market with
$\beta_0 = 0.1$, and {\bf (a)} long-term correlations ($\lambda =
0.01$) and {\bf (b)} short-time correlation ($\lambda = 0.05$).  Three
curves correspond to different transaction costs $\theta$: $0$ (solid
blue), $0.05$ (dashed green) and $0.15$ (dash-dotted red).  Vertical
black line indicates $\lambda$.}
\label{fig:Sharpe}
\end{figure}

It is instructive to consider the net risk adjusted P\&L of the
strategy, $\frac{\langle \dPNL_\infty \rangle - \langle
\TO_\infty\rangle}{\sqrt{v_\infty}}$, in which the mean turnover
$\langle \TO_\infty\rangle$ is included to account for transaction
costs.  In \ref{sec:turnover}, we derive the exact formula for the
mean daily turnover $\langle \TO_t \rangle$ and its stationary limit
$\langle \TO_\infty \rangle$.  For $\lambda \ll 1$, $\eta \ll 1$, and
linear transaction costs (i.e., $\alpha = 1$ in
Eq. (\ref{eq:TO_def})), Eq. (\ref{eq:TO}) becomes $\langle {\mathcal
T}_\infty \rangle \approx \frac{2}{\sqrt{\pi}} \theta \sqrt{\eta}$.
Using this approximate relation and approximations of
Eqs. (\ref{eq:dPNL_inf_exo}, \ref{eq:v_infty}) for $\lambda \ll 1$,
$\eta \ll 1$, and $\beta_0^2 \ll 1$, we obtain
\begin{equation}
\label{eq:NAPNL}
\frac{\langle \dPNL_\infty \rangle - \langle \TO_\infty\rangle}{\sqrt{v_\infty}} 
\approx \frac{\beta_0^2 \sqrt{2\eta} - \frac{2}{\sqrt{\pi}} \theta \sqrt{\eta} (\lambda+\eta)}{\sqrt{(\lambda+\eta)^2 + 2\beta_0^2(\lambda+\eta)}} .
\end{equation}
When there is no transaction cost (i.e., $\theta = 0$), this function
is maximized at $\eta_{\rm opt} = \lambda \sqrt{1 +
2\beta_0^2/\lambda}$, as illustrated on Fig. \ref{fig:Sharpe} (solid
curve).  For $\lambda = 0.01$ and $\beta_0 = 0.1$, the position of the
maximum is around $\lambda \sqrt{3}$, while the maximum level $0.8$ of
the {\it annualized} risk adjusted P\&L (given by Eq. (\ref{eq:NAPNL})
conventionally multiplied by $\sqrt{255}$) is a typical value for
systematic trading.  Interestingly, the optimal timescale of the
strategy is not equal to the timescale $\lambda$ of the market model
but it is enhanced by the factor $\sqrt{1+2\beta_0^2/\lambda}$ due to
auto-correlations of returns.  When transaction costs are included, an
explicit expression for the optimal timescale $\eta_{\rm opt}$ is too
lengthy.%
\footnote{
In fact, $\eta_{\rm opt} = \lambda z$, where $z$ is the positive root
of the cubic polynomial
$$\theta' z^3 + (c + \theta'(4c+3)) z^2 + 3\theta'(1+2c) z - (1+2c)(c-\theta') = 0 ,$$
which determines zeros of the derivative of Eq. (\ref{eq:NAPNL})
(here, $\theta' = \theta \sqrt{2/\pi}$ and $c = \beta_0^2/\lambda$).
Although an exact solution can be written, the formula is too lengthy
for further theoretical analysis.  In turn, this formula can be used
for numerical computation of the optimal timescale.}
As expected, an increase of the transaction cost $\theta$ reduces the
risk adjusted P\&L but also shifts the position of the maximum to
smaller $\eta$ in order to get smoother signal and thus reduce
transactions.  This behavior is illustrated in Fig. \ref{fig:Sharpe}.
Note that general formulas in \ref{sec:turnover} are also applicable
to nonlinear transaction costs.  Interestingly, the optimal timescale
depends on $\lambda$ and $\beta_0$ through the ratio
$\beta_0^2/\lambda$ which is of the order of unity.

Finally, the strategy is profitable only if the net risk adjusted P\&L
is positive, i.e., $\langle \dPNL_\infty\rangle \geq \langle
\TO_\infty \rangle$, from which one gets a simple condition on
transaction costs
\begin{equation}
\label{eq:theta}
\theta \leq \sqrt{\pi/2} ~ \frac{\beta_0^2}{\lambda + \eta} .
\end{equation}
The inequality (\ref{eq:theta}) can be seen as a limitation either on
the maximal transaction cost $\theta$, or on the minimal level of
auto-correlations $\beta_0$, or on the maximal timescale $\eta$ of the
strategy.

\subsection{Skewness and kurtosis}

\begin{figure}
\begin{center}
\includegraphics[width=67.5mm]{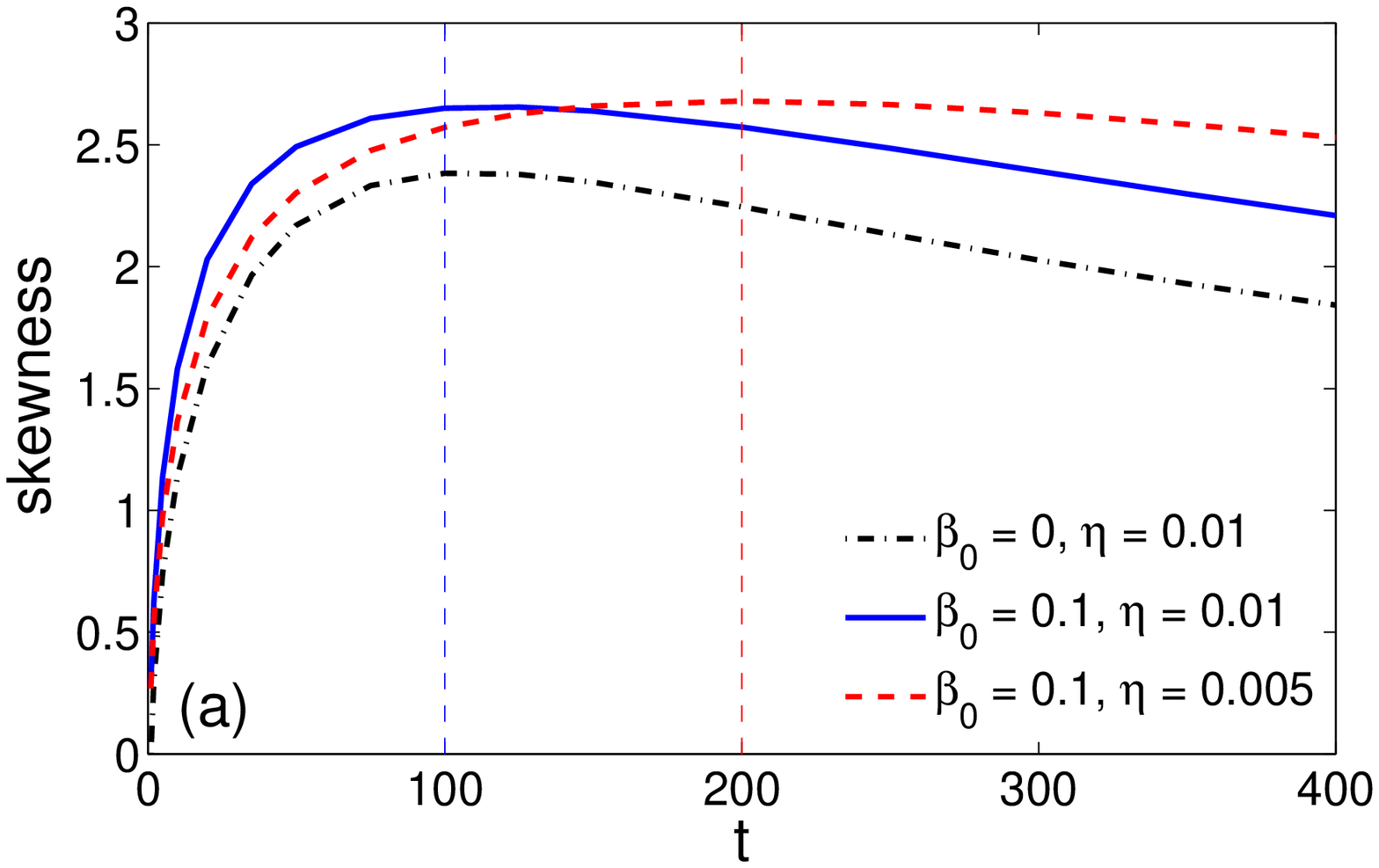}
\includegraphics[width=67.5mm]{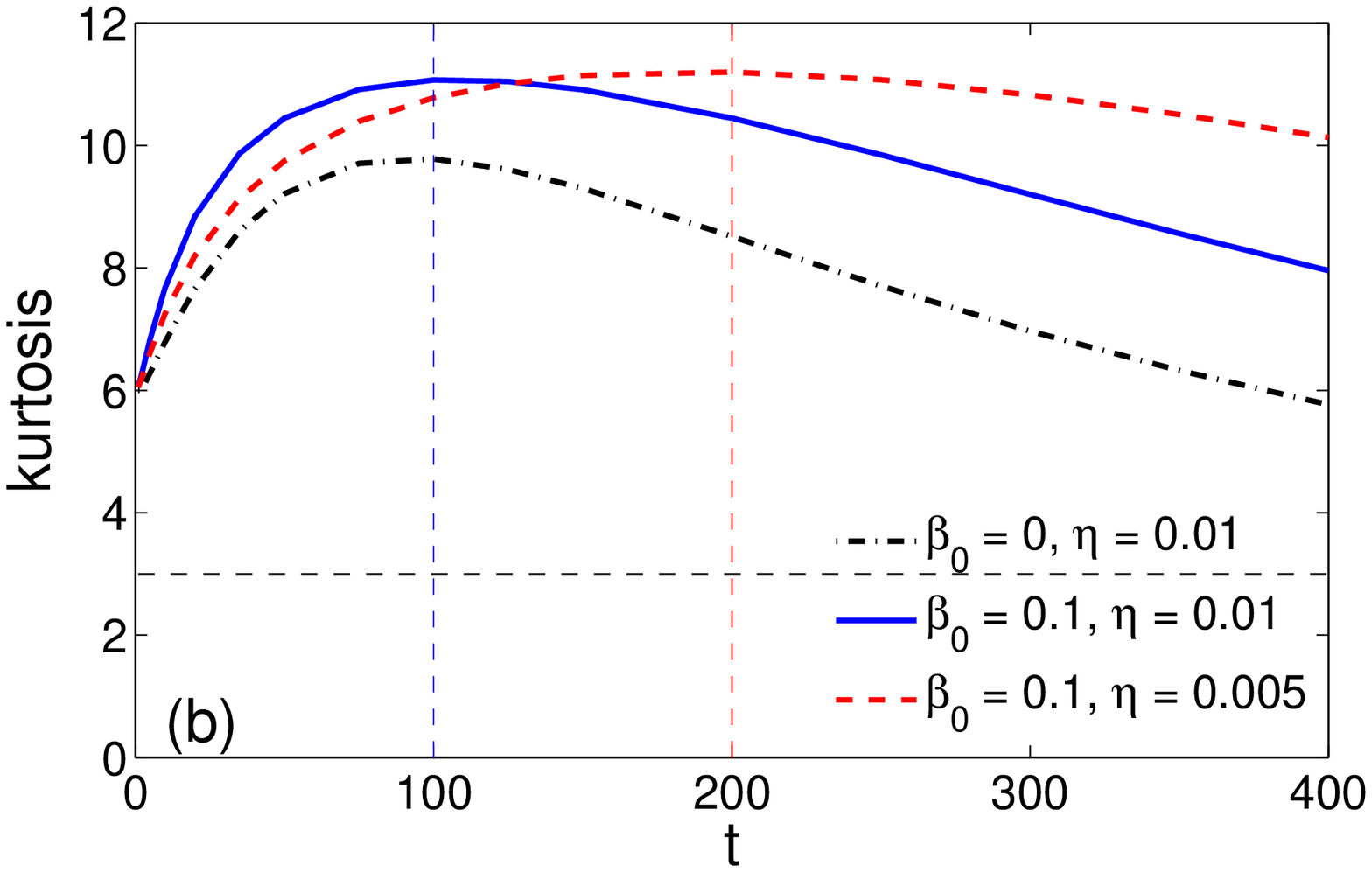}
\end{center}
\caption{
{\bf (a)} Skewness, $\kappa_3/\kappa_2^{3/2}$, and {\bf (b)} kurtosis,
$\kappa_4/\kappa_2^2$, of the cumulative P\&L, $\PNL_{t,t_0}$ (with
$t_0 = 200$ and two timescales, $\eta_1 = 0.01$ and $\eta_2 = 0.005$),
for auto-correlated returns ($\beta_0 = 0.1$ and $\lambda = 0.01$) and
independent returns ($\beta_0 = 0$).  Two vertical dashed lines
indicate $1/\eta_1$ and $1/\eta_2$, while horizontal dashed line
locates the level of kurtosis (equal to $3$) of Gaussian
distribution. }
\label{fig:skew}
\end{figure}

In principle, one can compute explicitly the other cumulant moments
and access skewness and kurtosis of the cumulative P\&L.  However,
these expressions become too lengthy for practical use.  In turn, the
general matrix formula (\ref{eq:kappa}) allows for rapid numerical
computation of these quantities.  Figure \ref{fig:skew} shows skewness
($\kappa_3/\kappa_2^{3/2}$) and kurtosis ($\kappa_4/\kappa_2^2$) of
the cumulative P\&L, $\PNL_{t,t_0}$, as functions of the lag time $t$.
Both quantities exhibit a maximum at $t \approx 1/\eta$, i.e., the
timescale of trend following strategy.  In other words, the strategy
induces auto-correlations of P\&Ls that are significant up to time
$1/\eta$ and then slowly decay.  In fact, if incremental P\&Ls,
$\dPNL_{t_0+1}$, ... , $\dPNL_{t_0+t}$, were independent and
identically distributed, the skewness and kurtosis of their sum,
$\PNL_{t,t_0}$, would decay as $1/\sqrt{t}$ and $1/t$, respectively.
We emphasize that this behavior of $\dPNL_t$ is induced by the trend
following strategy itself, irrespectively of auto-correlations of
returns.  This is confirmed by the fact that both skewness and
kurtosis behave similarly for independent and auto-correlated returns.

\subsection{Distribution of incremental P\&L}

According to Eq. (\ref{eq:dPNL}), the incremental P\&L, $\dPNL_{\t}$,
is the quadratic form defined by the symmetric matrix
$\M_{1-\eta}^{(1,\t-1)}$ from Eq. (\ref{eq:Mt}).  The probability
distribution of $\dPNL_\t$ can therefore be determined through the
inverse Fourier transform (\ref{eq:pz}).

\subsubsection{Independent returns}

We first consider the case of independent returns ($\beta_0 = 0$), for
which the covariance matrix is trivial: $\C = \I$.  In that case,
there are only two nonzero eigenvalues of the matrix $\M\C =
\M_{1-\eta}^{(1,\t-1)}$,
\begin{equation}
\mu_{\pm} = \pm \gamma ~ \sqrt{\frac{1-(1-\eta)^{2(\t-1)}}{1-(1-\eta)^2}} = \pm \sqrt{1 - (1-\eta)^{2(\t-1)}} ,
\end{equation}
where Eq. (\ref{eq:gamma}) was used in the last relation.  As a
consequence, the characteristic function of the incremental P\&L is
$\phi(k) = (1 + k^2 \mu_+^2)^{-1/2}$, from which the inverse Fourier
transform yields
\begin{equation}
\label{eq:pz_DPNL0}
p(z) = \frac{K_0(|z|/\mu_+)}{\pi \mu_+} ,
\end{equation}
where $K_0(x)$ is the modified Bessel function of the second kind.
For large $|z|$, the asymptotic behavior is
\begin{equation}
p(z) \simeq \frac{\exp(-|z|/\mu_+)}{\sqrt{2\pi \mu_+ |z|}} \bigl(1 + O(1/|z|)\bigr)  \qquad (|z|\to\infty).
\end{equation}
Note that $\mu_+^2$ is the variance of the incremental P\&L.  The
skewness and kurtosis are $0$ and $6$, respectively (see
Fig. \ref{fig:skew}, on which an incremental P\&L corresponds to
$t=1$).  In the stationary limit $t_0\to \infty$, one gets $\mu_+ =
1$.

Figure \ref{fig:distrib_dPNL} compares the probability density $p(z)$
of the incremental P\&L, $\dPNL_\t$, with the Gaussian density
$(2\pi)^{-1/2} \exp(-z^2/2)$ of a single return $r_\t$ (with the unit
variance).  Although the mean and variance of these two distributions
are identical, their overall behaviors are drastically different.  The
incremental P\&L is peaked at $0$ (in fact, $K_0(z)$ logarithmically
diverges at $0$), while the tail decay is much slower than for
returns.  This transformation from a Gaussian density to $p(z)$ is the
effect of a trend following strategy.

\subsubsection{Auto-correlated returns}

\begin{figure}
\begin{center}
\includegraphics[width=67.5mm]{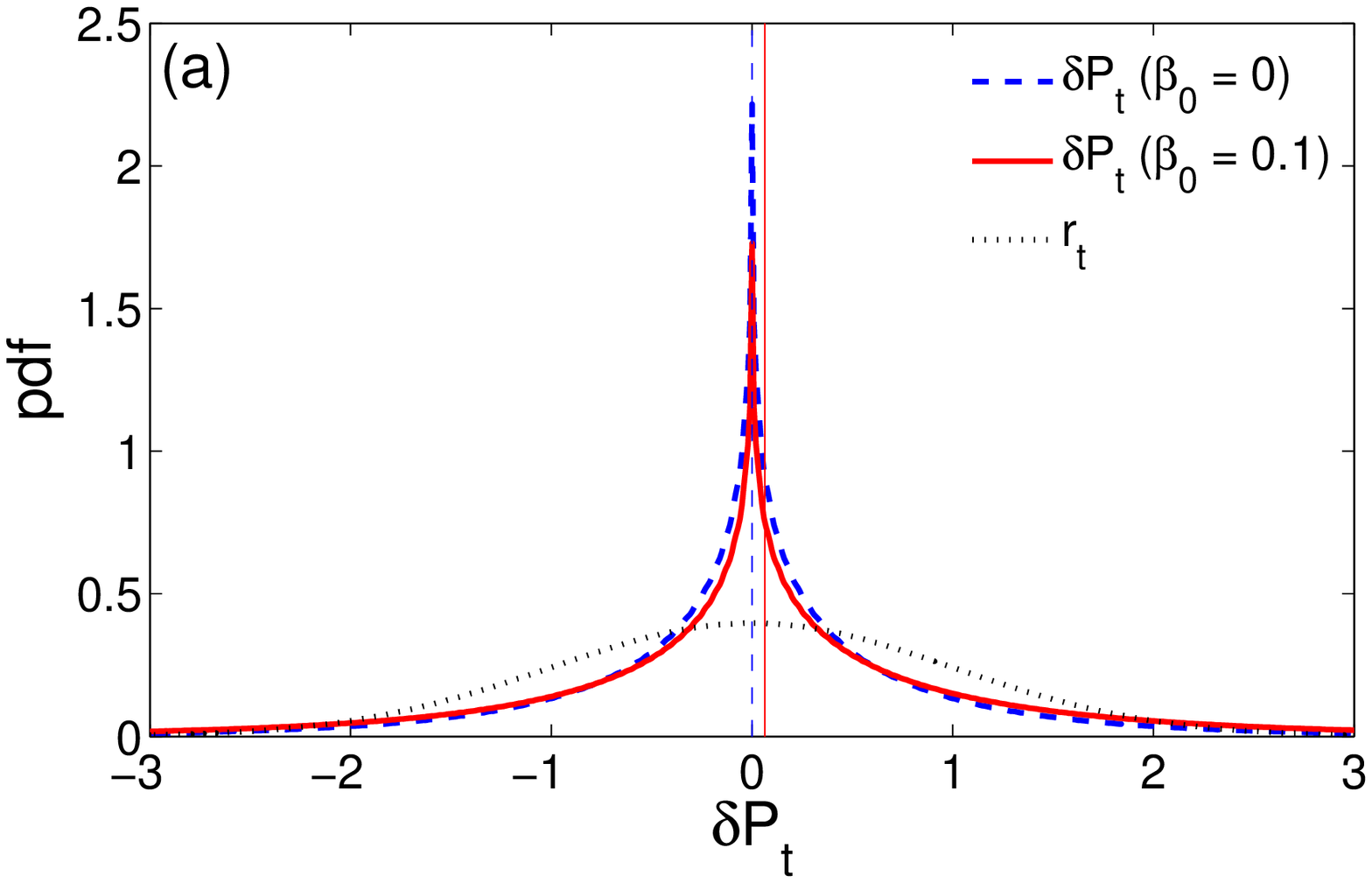}
\includegraphics[width=67.5mm]{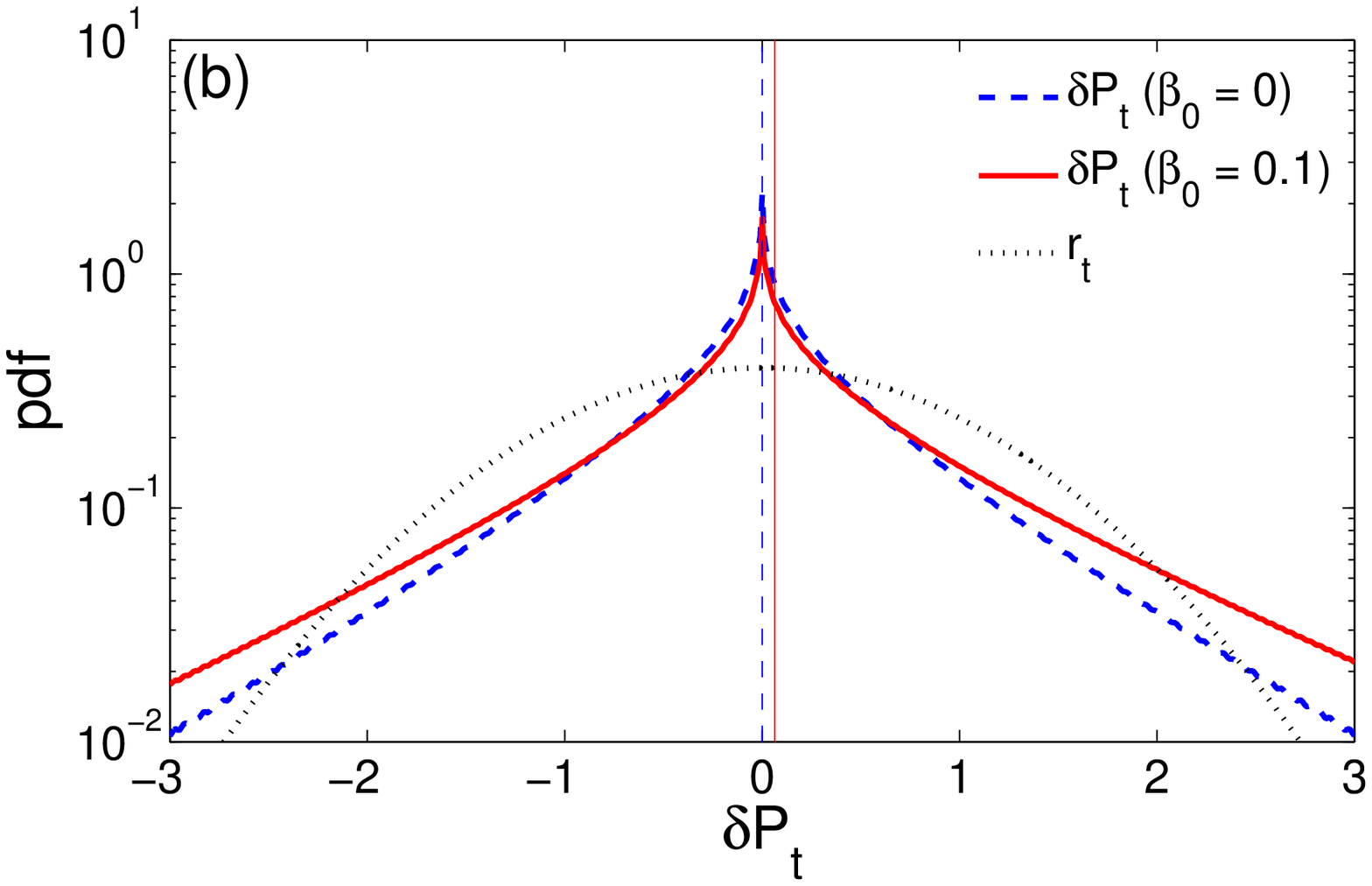}
\end{center}
\caption{
Probability distribution of the incremental P\&L, $\dPNL_\t$ ($\t =
200$ and $\eta = 0.01$), for independent returns (dashed blue line,
$\beta_0 = 0$, Eq. (\ref{eq:pz_DPNL0})) and for auto-correlated
returns (solid red line, $\beta_0 = 0.1$, $\lambda = 0.01$): {\bf (a)}
linear scale, {\bf (b)} semi-logarithmic scale.  The Gaussian
distribution of returns $r_\t$ is plotted for comparison (dotted black
line).  Vertical lines indicate the mean value $\langle
\dPNL_\t\rangle$ for both cases (note that the variance of $\dPNL_\t$
is equal to $1$ for independent returns and to $1.01$ for
auto-correlated returns). }
\label{fig:distrib_dPNL}
\end{figure}

For auto-correlated returns ($\beta_0 \ne 0$), the diagonalization of
the matrix $\M\C$ and computation of the probability density $p(z)$
can be performed numerically.  As illustrated on
Fig. \ref{fig:distrib_dPNL}, small auto-correlations of returns
($\beta_0 = 0.1$) slightly modify the probability distribution
(\ref{eq:pz_DPNL0}) by shifting the mean $\langle \dPNL_t\rangle$ from
zero to a small positive value and by increasing the probability of
extreme values of $\dPNL_t$ (both positive and negative tails).


\subsection{Distribution of cumulative P\&L}

\begin{figure}
\begin{center}
\includegraphics[width=67.5mm]{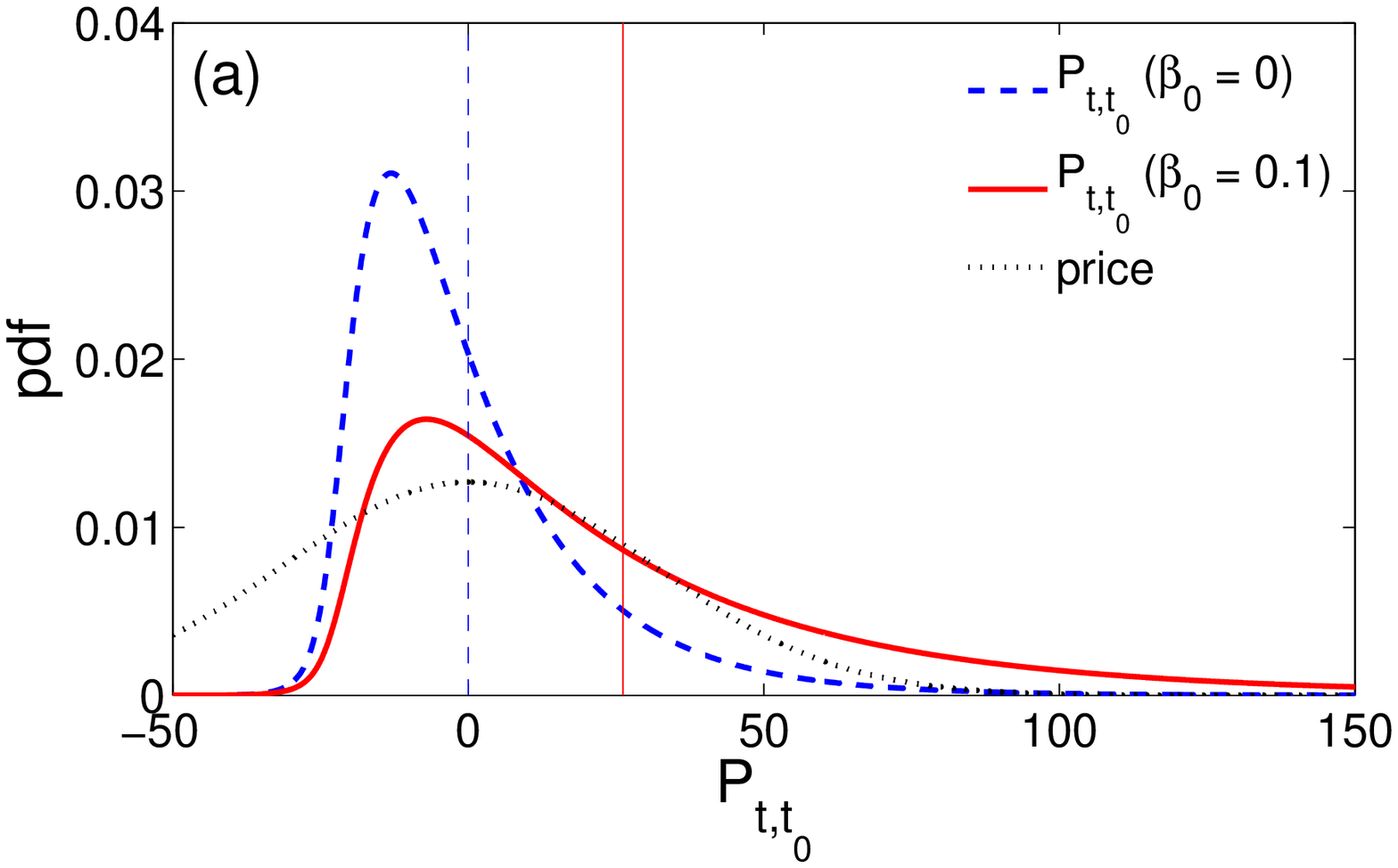}
\includegraphics[width=67.5mm]{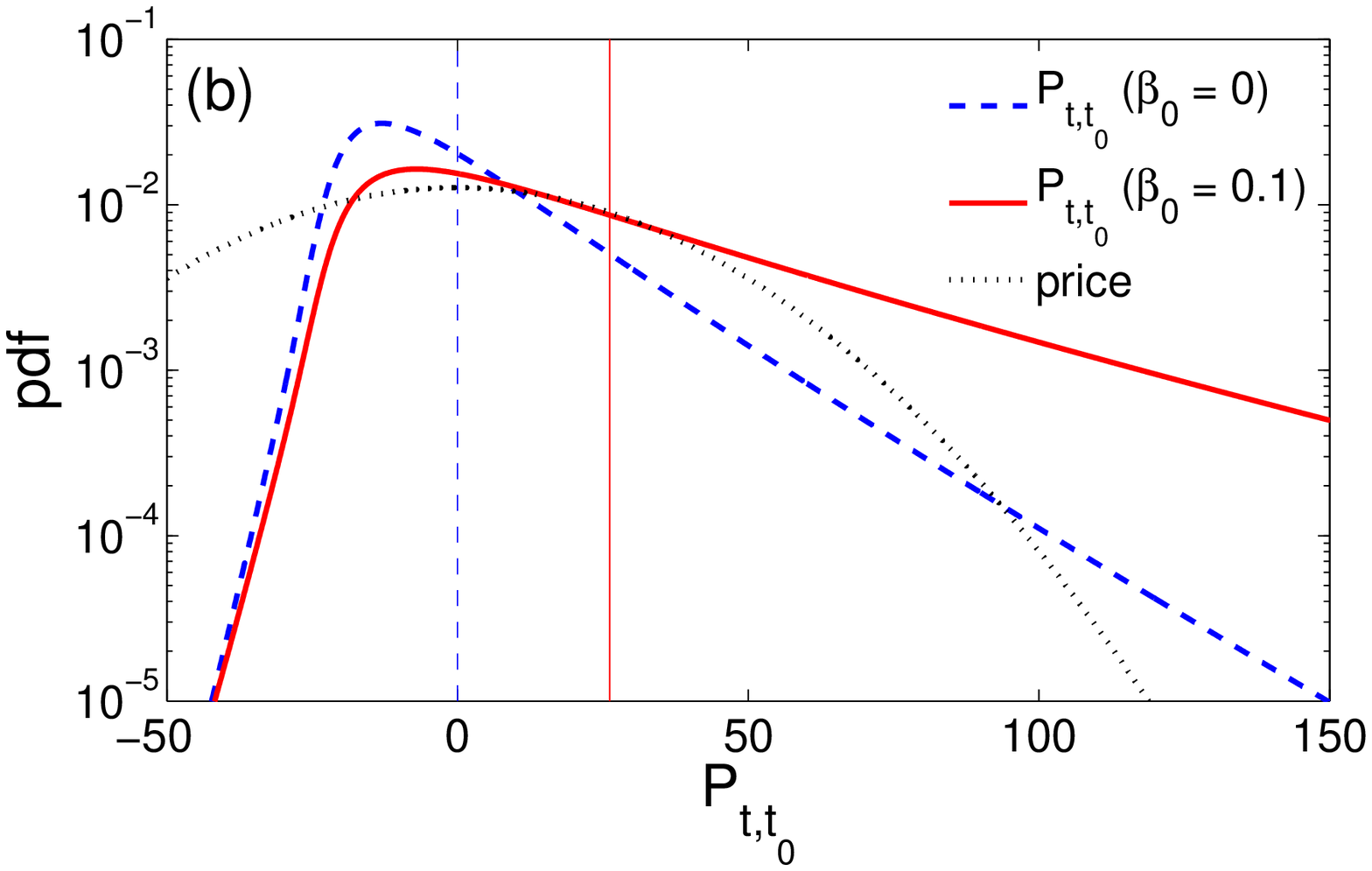}
\end{center}
\caption{
Probability distribution of the cumulative P\&L, $\PNL_{t,t_0}$ ($t =
300$, $t_0 = 200$, and $\eta = 0.01$), for independent returns (dashed
blue line, $\beta_0 = 0$) and for auto-correlated returns (solid red
line, $\beta_0 = 0.1$, $\lambda = 0.01$): {\bf (a)} linear scale, {\bf
(b)} semi-logarithmic scale.  The Gaussian distribution of price
variation, $r_{t_0+1} + ... + r_{t_0+t}$, is plotted for comparison
(dotted black line).  Vertical lines indicate the mean value $\langle
\PNL_{t,t_0}\rangle$ for both cases. }
\label{fig:distrib_PNL}
\end{figure}

The probability distribution of the cumulative P\&L, $\PNL_{t,t_0}$,
can be obtained numerically through the inverse Fourier transform
(\ref{eq:pz}).  Figure \ref{fig:distrib_PNL} shows the probability
density $p(z)$ of $\PNL_{t,t_0}$ for independent returns ($\beta_0 =
0$) and for auto-correlated returns (with $\beta_0 = 0.1$ and $\lambda
= 0.01$).  The initiation period of $t_0 = 200$ points was ignored to
achieve stationary properties.  In sharp contrast to symmetric (or
almost symmetric) distributions of the incremental P\&L from
Fig. \ref{fig:distrib_dPNL}, the distribution of the cumulative P\&L
is strongly skewed and asymmetric, even for independent returns, in
agreement with earlier observations \cite{Potters05,Martin12}.  The
most probable P\&L is negative, while the mean $\langle
\PNL_{t,t_0}\rangle$ is nonnegative (it is $0$ for $\beta_0 = 0$ and
strictly positive for $\beta_0 \ne 0$).  The positive mean P\&L for
$\beta_0 \ne 0$ is ensured by relatively large probability for getting
large positive profits.  This result agrees with earlier observations
that trend followers experience often small losses, waiting for a
trend that may lead to considerable profits \cite{Potters05}.  We
emphasize again that skewness of $\PNL_{t,t_0}$ emerges due to the
trading strategy itself, irrespectively of auto-correlations of
returns.

\subsection{Quantiles}

Inspecting the distribution on Fig. \ref{fig:distrib_PNL}, one can
clearly observe an exponential decay of both positive and negative
tails, in agreement with the expected asymptotic behavior
(\ref{eq:pz_asympt}).  Importantly, the decay of the probability of
negative P\&Ls is much steeper than that of positive P\&Ls.  These
extreme events can be characterized by quantiles.  For this purpose,
one first integrates the density $p(z)$ to get the cumulative
probability distribution $F(z)$ and then solves the equation $F(z_q) =
q$ with $0 < q < 1$ that determines the $q$-quantile $z_q$ of the
distribution.  In a first approximation, power law corrections in
Eq. (\ref{eq:pz_asympt}) can be ignored (by setting $\nu_\pm = 0$) so
that
\begin{equation*}
\begin{split}
F(z) & \approx A_- |\mu_-| \exp(-z/\mu_-) \hskip 5.5mm (z\to -\infty) , \\
1-F(z) & \approx A_+ \mu_+ \exp(-z/\mu_+) \qquad (z\to \infty) . \\
\end{split}
\end{equation*}
Extreme negative values of the cumulative P\&L correspond to the limit
$q\to 0$ for which the equation $F(z_q) = q$ can be approximately
solved as
\begin{equation}
z_q \approx -|\mu_-| \ln\left(\frac{A_-|\mu_-|}{q}\right) \qquad (q\to 0) .
\end{equation}
As expected, the behavior of the small $q$-quantile $z_q$ is
essentially determined by the smallest eigenvalue $\mu_-$ of the
matrix $\M\C$.  In turn, extreme positive values of P\&L correspond to
the limit $q\to 1$ for which
\begin{equation}
z_q \approx \mu_+ \ln\left(\frac{A_+ \mu_+}{1-q}\right) \qquad (q\to 1) .
\end{equation}
The large $q$-quantile is therefore mainly determined by the largest
eigenvalue $\mu_+$ of the matrix $\M\C$.  

\begin{figure}
\begin{center}
\includegraphics[width=67.5mm]{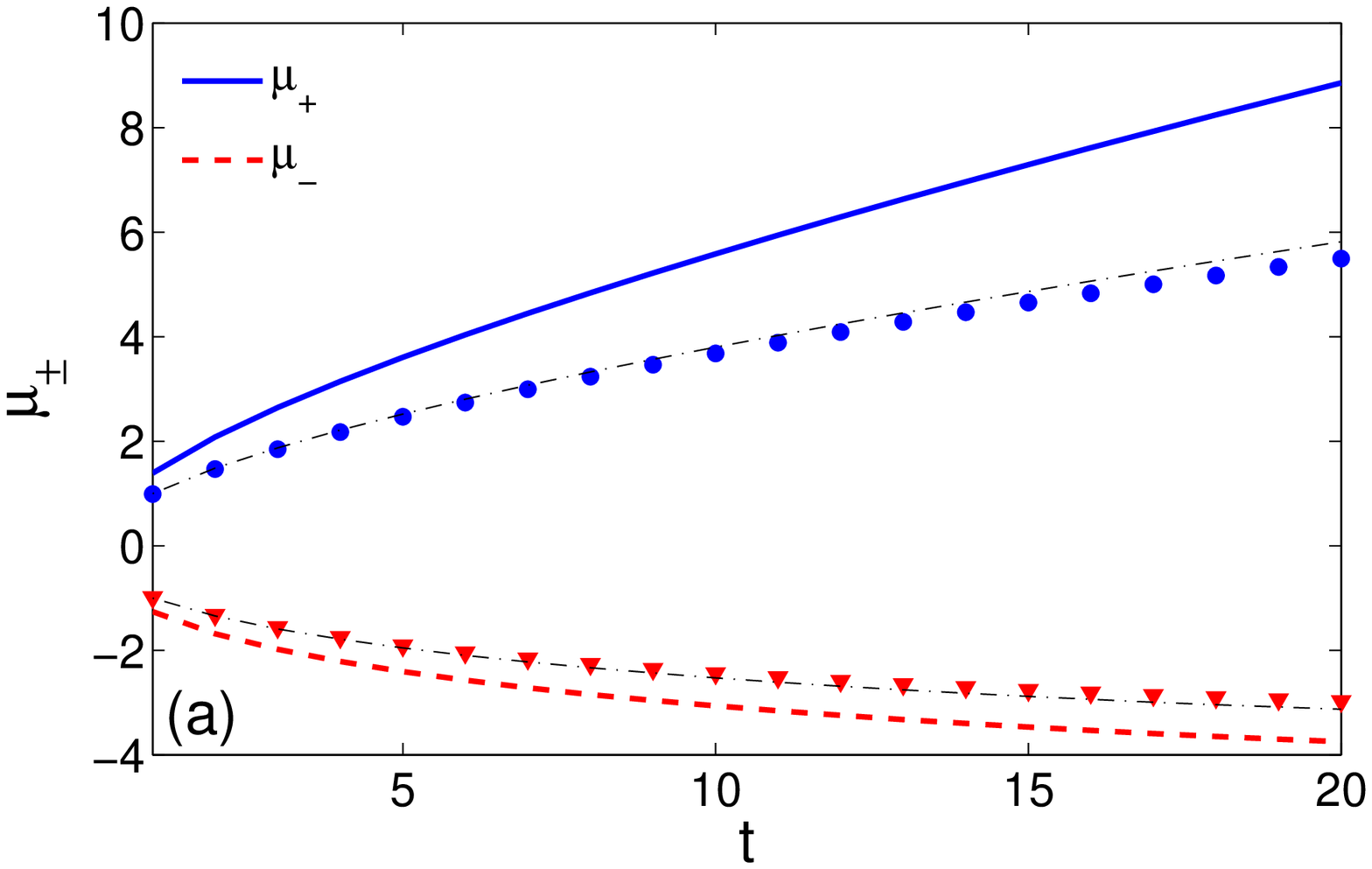}
\includegraphics[width=67.5mm]{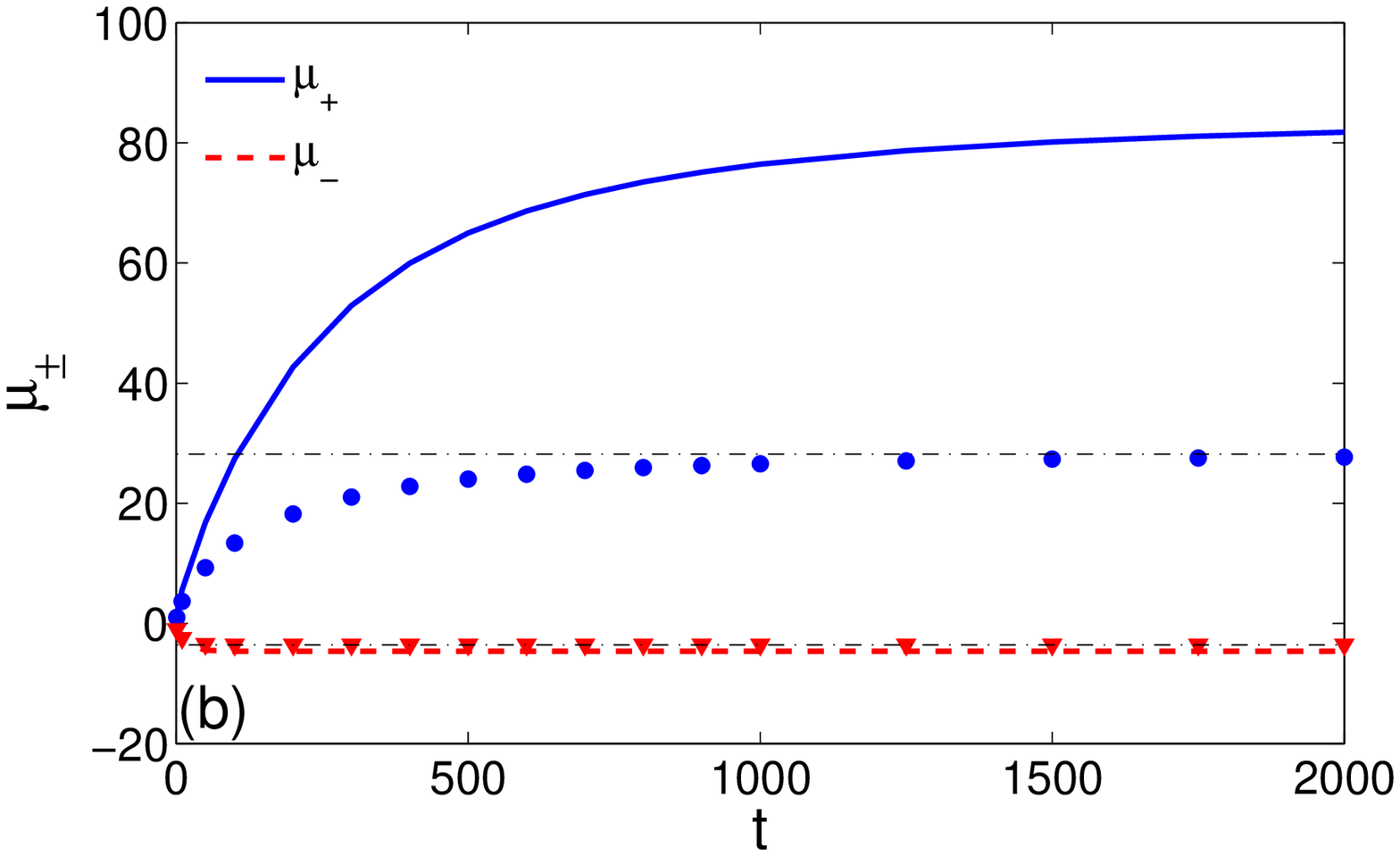}
\end{center}
\caption{
The largest and the smallest eigenvalues of the matrix
$\M_{1-\eta}^{(t,t_0)}\C$ (with $t_0 = 200$, $\eta = 0.01$), for
independent returns (symbols) and for auto-correlated returns (lines,
$\beta_0 = 0.1$, $\lambda = 0.01$).  {\bf (a)} Short-time behavior.
Black dash-dotted lines indicate the asymptotic behavior
(\ref{eq:asymp_short}). {\bf (b)} Long-time behavior.  Horizontal
black dash-dotted lines indicate the asymptotic values
$\mu_\pm^\infty$ from Eqs. (\ref{eq:eigen_mup1},
\ref{eq:eigen_mum1}).}
\label{fig:eigen}
\end{figure}

The behavior of the smallest and the largest eigenvalues $\mu_\pm$ of
the matrix $\M_{1-\eta}^{(t,t_0)}\C$ for the cumulative P\&L is
illustrated on Fig. \ref{fig:eigen}.  The largest eigenvalue $\mu_+$
grows with time $t$ and slowly approaches a constant value at long
$t$.  In turn, the smallest eigenvalue $\mu_-$ decreases and
approaches a constant value much faster.  For independent returns, we
compute in \ref{sec:spectrum_long} the asymptotic values
$\mu_\pm^\infty$:
\begin{eqnarray}
\label{eq:eigen_mup1}
\mu_+^\infty & = & \frac{2\sqrt{\eta(2-\eta)}}{\eta} \approx \frac{\sqrt{8}}{\sqrt{\eta}} , \\
\label{eq:eigen_mum1}
\mu_-^\infty & = & - \frac{\sqrt{\eta(2-\eta)}}{2\eta(1-\eta)(2-\eta)} \approx - \frac{1}{\sqrt{8\eta}}. 
\end{eqnarray}
These values are shown on Fig. \ref{fig:eigen}b by horizontal
dash-dotted lines.  One can see that $\mu_+^\infty$ is 8 times larger
than $|\mu_-^\infty|$.  Most importantly, Eq. (\ref{eq:eigen_mum1})
turns out to be an accurate approximation for the smallest eigenvalue
$\mu_-$ even for auto-correlated returns.  In other words, extreme
negative P\&Ls weakly depend on the market features (here, $\beta_0$
and $\lambda$) and are mainly determined by the trend following
strategy (timescale $\eta$).  In turn, the largest eigenvalue $\mu_+$
for auto-correlated returns may attain much larger values than
$\mu_+^\infty$ from Eq. (\ref{eq:eigen_mup1}).  In other words, the
presence of trends due to auto-correlations of returns increases
$\mu_+$ and thus enhances the probability of extreme positive P\&Ls.
In contrast to $\mu_-$, the largest eigenvalue $\mu_+$ is sensitive to
the market features.

We also analyzed the behavior of the largest and the smallest
eigenvalues $\mu_\pm$ in the opposite case of short times $t$.  For
independent returns, we compute in \cite{SM} the eigenvalues of the
matrix $\M_{1-\eta}^{(t,t_0)}\C$ for $t = 2,3,4$ and $t_0\to\infty$.
These explicit results suggest the conjectural asymptotic relation
\begin{equation}
\label{eq:asymp_short}
\mu_\pm \simeq \pm \sqrt{t} + (t-1) \sqrt{\eta/2} + O(\eta) ,
\end{equation}
which is applicable for small $\eta$ and moderate values of $t$ (see
Fig. \ref{fig:eigen}a).  At short times, the small $q$-quantile can be
approximated as
\begin{equation}
\label{eq:quantile_q0}
z_q \approx -\sqrt{t} \biggl(1 - \frac{t-1}{\sqrt{2t}} \sqrt{\eta} \biggr) \ln\left(\frac{A_- \sqrt{t}}{q}\right) \qquad (q\to 0) .
\end{equation}

This behavior can be compared to the quantile of price variation over
the time $t$, $r_{t_0+1} + ... + r_{t_0 + t}$.  For independent
returns, this is a Gaussian variable with mean zero and variance $t$,
independently of the initiation period duration $t_0$.  The cumulative
probability distribution is $\frac12(1 + \erf(z/\sqrt{2t}))$, from
which the $q$-quantile is given by the inverse error function:
\begin{equation}
\label{eq:quantile_Gauss}
z_q^0 = 2\sqrt{t}~ \erf^{-1}(2q-1) .
\end{equation}
One can see that this quantile grows as $\sqrt{t}$, while
$z_q^0/\sqrt{t}$ is constant.  In turn, the quantile for the P\&L,
even after normalization by $\sqrt{t}$, exhibits a power law increase
according to Eq. (\ref{eq:quantile_q0}).

\subsubsection{Auto-correlated returns}

For auto-correlated returns ($\beta_0 \ne 0$), the quantiles were
computed numerically by solving the equation $F(z_q) = q$, in which
the cumulative probability distribution $F(z)$ was found by
integrating the probability density $p(z)$.

\begin{figure}
\begin{center}
\includegraphics[width=67.5mm]{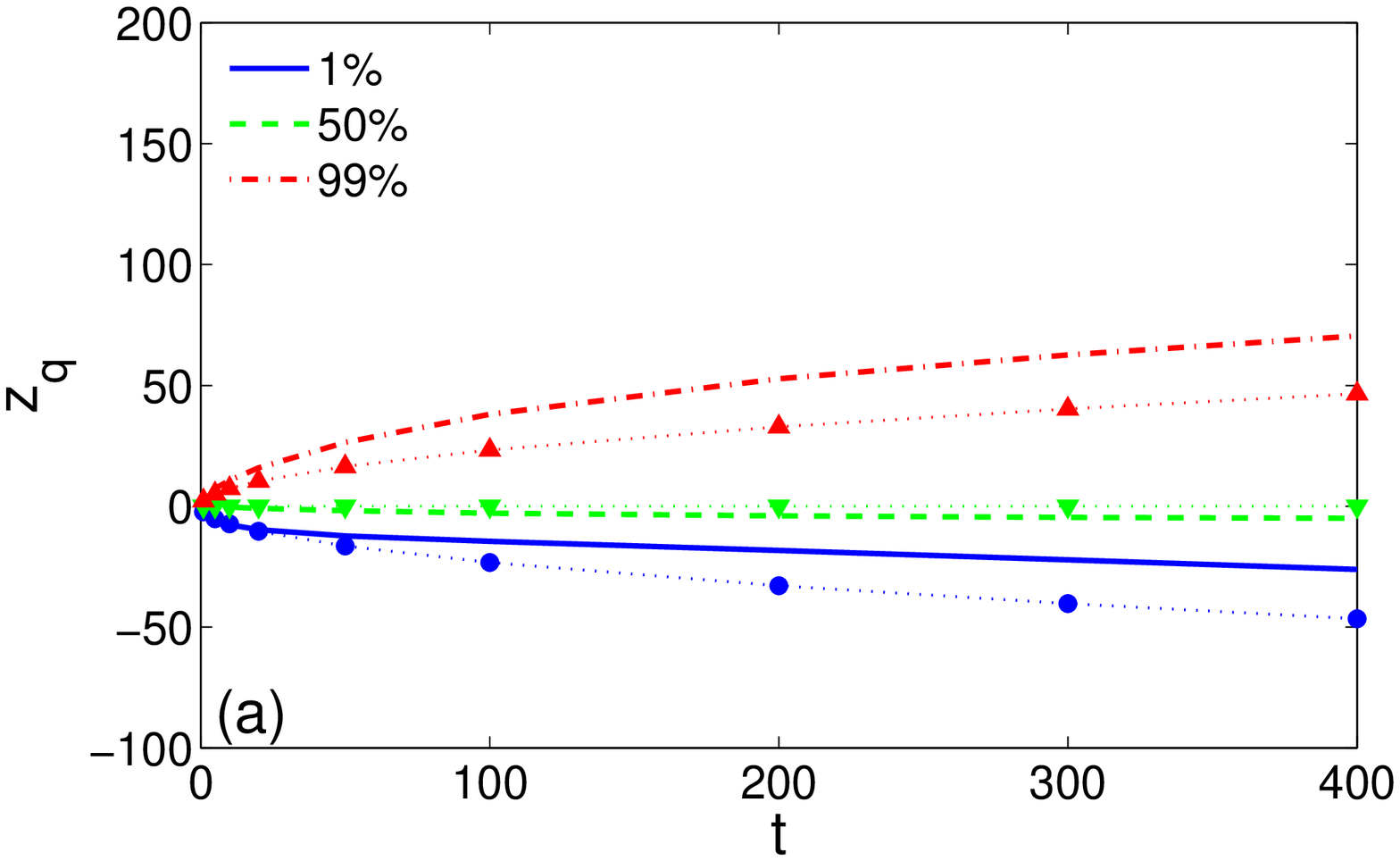}
\includegraphics[width=67.5mm]{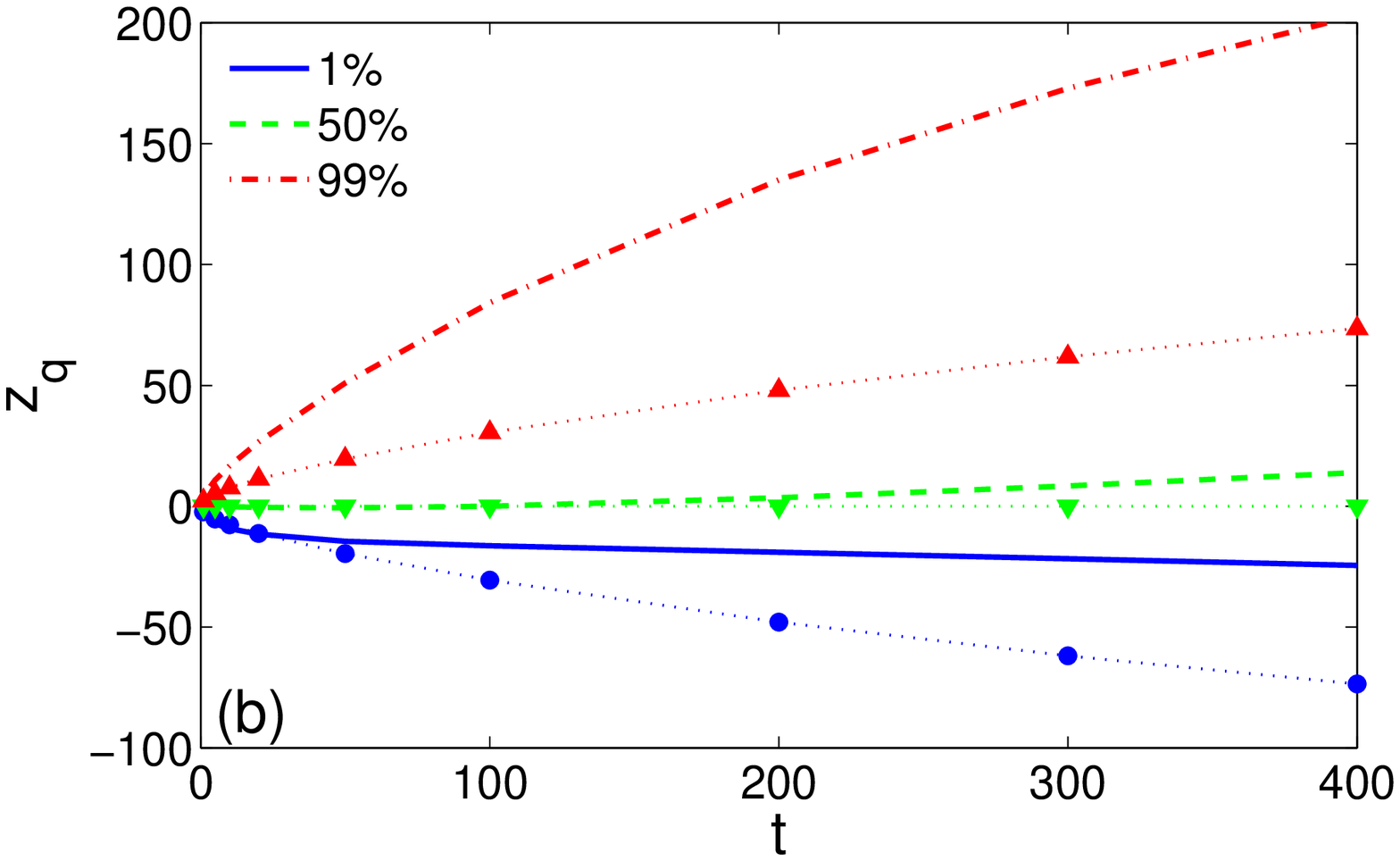}
\includegraphics[width=67.5mm]{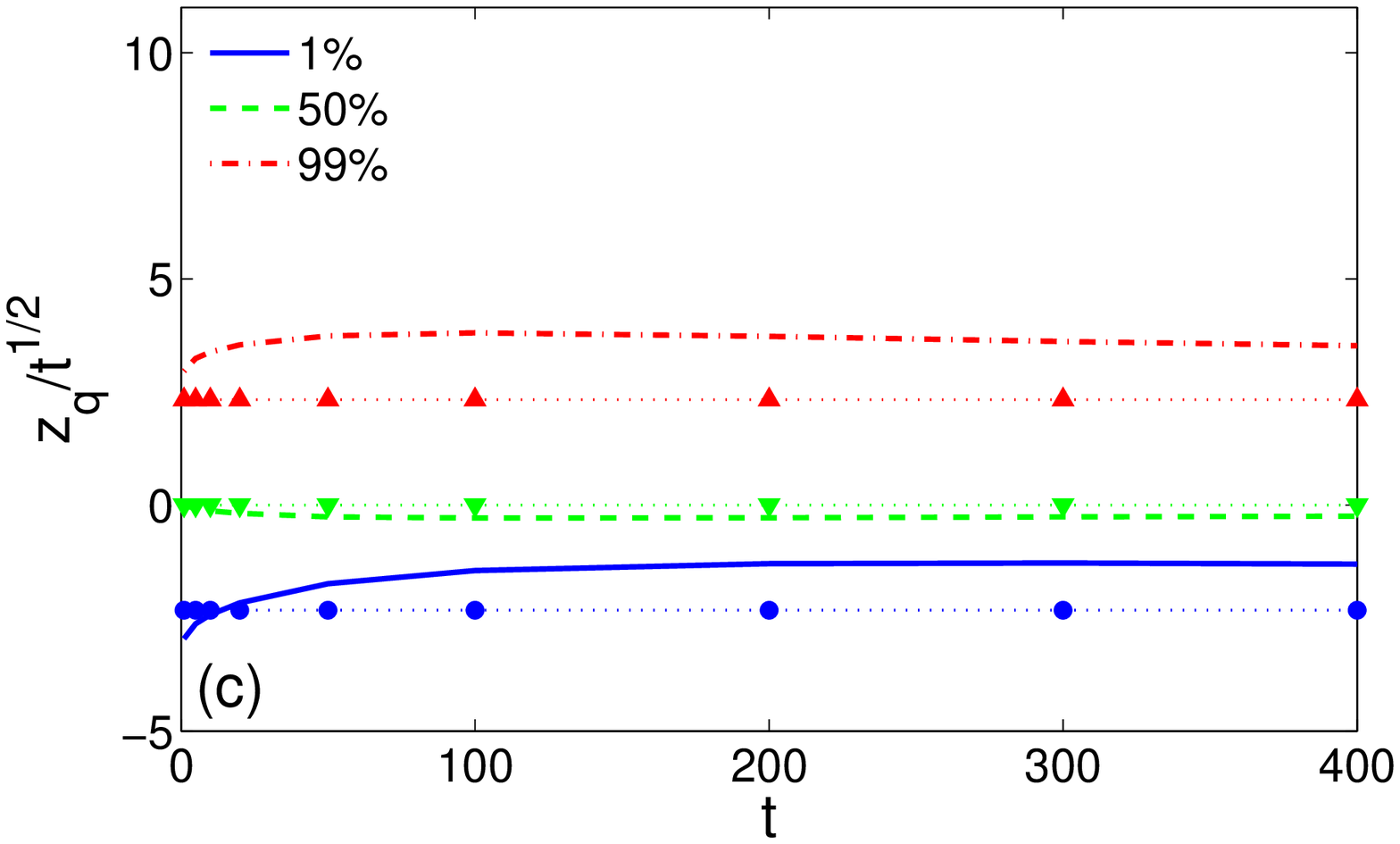}
\includegraphics[width=67.5mm]{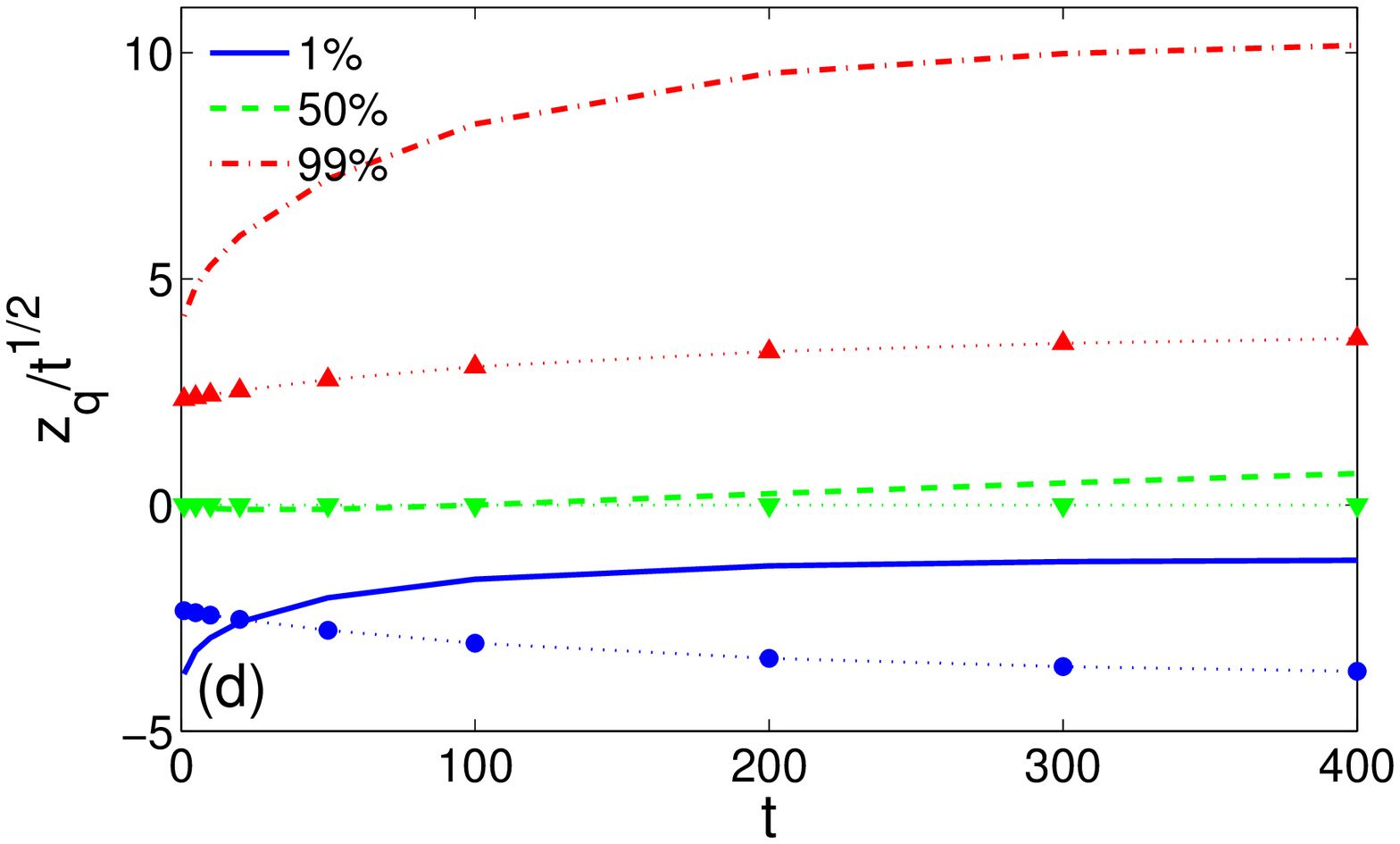}
\end{center}
\caption{
Quantiles $z_{0.01}$, $z_{0.5}$ and $z_{0.99}$ ($1\%$, $50\%$, and
$99\%$) of the cumulative P\&L, $\PNL_{t,t_0}$ as functions of time
$t$, with $\eta = 0.01$ and $t_0 = 200$: {\bf (a)} for independent
returns ($\beta_0 = 0$), and {\bf (b)} auto-correlated returns
($\beta_0 = 0.1$, $\lambda = 0.01$).  The same quantiles of price
variation over time $t$, $r_{t_0+1} + ... + r_{t_0+t}$, are shown by
symbols.  Since Gaussian quantiles grow as $\sqrt{t}$, the ratio
$z_q/\sqrt{t}$ is also plotted for independent returns {\bf (c)} and
for auto-correlated returns {\bf (d)}, with the same parameters.}
\label{fig:quantiles}
\end{figure}

Figure \ref{fig:quantiles} illustrates the behavior of quantiles for
the cumulative P\&L, $\PNL_{t,t_0}$, and for price variation over the
same time $t$, $r_{t_0+1} + ... + r_{t_0+t}$.  For independent returns
($\beta_0 = 0$), the Gaussian quantile grows as $\sqrt{t}$ according
to Eq. (\ref{eq:quantile_Gauss}).  In turn, the quantiles for
$\PNL_{t,t_0}$ exhibit quite different behavior showing a strong
asymmetry between positive and negative values.  This asymmetry is
further enhanced by auto-correlations of returns.  The most
interesting feature is the behavior of the renormalized quantile
$z_q/\sqrt{t}$ for small $q = 0.01$ illustrated on
Fig. \ref{fig:quantiles}d.  At short time $t$, the negative values of
this quantile are smaller for P\&L than for price variation.  In other
words, a trend following strategy may lead to more significant losses
than one could naively anticipate from a Gaussian distribution of
price variations.  At larger times, the situation changes to the
opposite: the quantile for the P\&L exceeds that for price variation
quite significantly.  In other words, the trend following strategy
ensures {\it smaller} losses at {\it longer} times, even in the
absence of trends (when $\beta_0 = 0$, see
Fig. \ref{fig:quantiles}a,c).

One can also observe many similarities between
Fig. \ref{fig:DJ_quantiles}a,b (Dow Jones) and
Fig. \ref{fig:quantiles}a,b (present model).  In particular, the $1\%$
quantiles in both cases are close to each other, confirming that
quantiles for extreme negative P\&Ls weakly depend on market features
and may be well approximated even by a simple model.  In turn, the
$99\%$ quantiles on Fig. \ref{fig:DJ_quantiles}b and
Fig. \ref{fig:quantiles}b are not so close, i.e., quantiles for
extreme positive P\&Ls are sensitive to market features.

\section{Conclusion}
\label{sec:discussion}

In this paper, we investigated how price variations of a stock are
transformed into profits and losses of a trend following strategy.  We
started by deriving simple formulas for the mean and variance of the
P\&L, as well as the mean turnover of the strategy.  The explicit
expression for the net annualized risk adjusted P\&L allowed us to
analyze the profitability of trend following strategies in the
presence of auto-correlations and transaction costs, and their
sensitivity to the choice of parameters.  We next proceeded in
computing explicitly the probability distribution of P\&L and
investigating its asymptotic behavior.  The theoretical analysis was
mainly done for independent returns and confronted to numerical
results for auto-correlated returns.  Although the model of correlated
returns was over-simplified, it allowed us to illustrate the basic
features and mechanisms of a trend following strategy.  Moreover, the
general matrix formulas provided at the beginning of
Sec. \ref{sec:PNL} are applicable to arbitrary Gaussian models.

It is worth emphasizing that quantitative results of this study are
model dependent.  For instance, we analyzed the asymptotic behavior of
the probability density $p(z)$ for extreme losses and showed its
exponential decay with the rate $\mu_-^\infty \approx
-1/\sqrt{8\eta}$.  An exponential decay is a universal feature for
quadratic forms of Gaussian vectors.  However, the dependence of the
smallest eigenvalue $\mu_-$ on the parameter $\eta$ was only derived
for the studied trend following strategy.  Moreover, the unknown
prefactor $A_-$ in the asymptotic formula (\ref{eq:pz_asympt}) and
related quantiles may strongly depend on other parameters, rendering
quantitative estimates of quantiles model dependent.  Finally, the
asymptotic exponential decay of $p(z)$ may settle at extremely large
values of $|z|$, at which the probability density is negligible and out
of practical interest.
%
%
At the same time, qualitative conclusions of the study are expected to
be general.  In fact, a trend following strategy strongly modifies
probabilistic properties of price time series, yielding skewed
asymmetric distributions, with often small losses and less frequent
high profits.  The probability of extreme losses decays much faster
than the probability of extreme profits.  Moreover, the occurence of
extreme losses is more influenced by the trend following strategy than
by the market itself.  In turn, the occurence of extreme profits
depends on both the strategy and the market.  We showed that the usual
Gaussian paradigm may lead to erroneous conclusions about trend
following strategies.  For instance, at short times, trend following
strategies admit larger losses than one may anticipate from standard
Gaussian estimates.  This is an important message for systematic
traders and risk analysts.

The present analysis can be extended to arbitrary Gaussian models of
returns and to multiple correlated stocks.  The practical advantage of
choosing {\it linear} equation (\ref{eq:St_prop}) is that the signal
from EMAs of individual stocks is simply the sum of the related
signals.  As a consequence, the P\&L of a portfolio is again a
quadratic form of Gaussian vectors for which general matrix formulas
at the beginning of Sec. \ref{sec:PNL} are still applicable.  One can
therefore study the role of inter-stock correlations which may
significantly improve risk control of trend following strategies.

\appendix
\section{Mean turnover of trend following strategy}
\label{sec:turnover}

Accounting for transaction costs is important for a comprehensive
analysis of trading strategies.  We define the daily turnover of the
trend following strategy as
\begin{equation}
\label{eq:TO_def}
{\mathcal T}_t = \theta ~|s_t - s_{t-1}|^{\alpha} ,
\end{equation}
where $\theta$ represents transaction cost, and $\alpha$ is an
appropriate exponent (typically $\alpha = 1$ or $\alpha = 2$).  The
mean turnover can be evaluated by using the identity
\begin{equation}
\langle f\bigl((\r^\T \a)\bigr)\rangle = \int\limits_{-\infty}^\infty dz ~ f(z) ~ 
\frac{1}{\sqrt{2\pi(\a^\T \C \a)}} \exp\left(-\frac{z^2}{2(\a^\T \C \a)}\right) ,
\end{equation}
where $f(z)$ is a continuous function of the scalar product $(\r^\T
\a)$, $\r$ is a Gaussian vector with mean zero and covariance matrix
$\C$, and $\a$ is a fixed vector.  Setting $f(z) = |z|^\alpha$ and
$a_j = \gamma \bigl[(\A_{1-\eta})_{t,j} -
(\A_{1-\eta})_{t-1,j}\bigr]$, one gets
\begin{equation}
\begin{split}
\langle {\mathcal T}_t \rangle & = \theta ~\frac{\Gamma(\frac{1+\alpha}{2})}{\sqrt{\pi}}  \bigl[2(\a^\T \C \a)\bigr]^{\alpha/2}  
 = \theta ~\frac{\Gamma(\frac{1+\alpha}{2})}{\sqrt{\pi}}  (2\gamma^2)^{\alpha/2} \\
& \times  \left( \bigl[\A_p\C \A_p^\T\bigr]_{t,t}
- 2\bigl[\A_p\C \A_p^\T\bigr]_{t,t-1} + \bigl[\A_p\C \A_p^\T\bigr]_{t-1,t-1} \right)^{\alpha/2} , \\
\end{split}
\end{equation}
where $\Gamma(z)$ is Gamma function.  Using Eq. (\ref{eq:C1}), one
obtains explicitly
\begin{equation}
\begin{split}
& \langle {\mathcal T}_t \rangle  = \theta ~\frac{\Gamma(\frac{1+\alpha}{2})}{\sqrt{\pi}}  (2\gamma^2)^{\alpha/2}
 \biggl( \frac{2-(1-p) p^{2t-4}}{1+p} + \frac{\beta_0^2 (1-q^2)}{(1-pq)(p-q)} 
 \biggl[\frac{2(p-q)}{(1+p)(1+q)} \\
& - \frac{(1-p)p^{2t-3}}{1+p} + \frac{(1-q)q^{2t-3}}{1+q} - \frac{(p^{t-1}-p^{t-2} - q^{t-1}+q^{t-2})^2}{p-q}\biggr] \biggr)^{\alpha/2}.  \\
\end{split}
\end{equation}
In the special case $p = q$, one gets
\begin{equation}
\begin{split}
\langle {\mathcal T}_t \rangle & = \theta ~\frac{\Gamma(\frac{1+\alpha}{2})}{\sqrt{\pi}}  (2\gamma^2)^{\alpha/2} \biggl( \frac{2-(1-q) q^{2t-4}}{1+q} \\
& + \frac{\beta_0^2}{(1+q)^2} \biggl[2 - q^{2t-4} \biggl(1 + \bigl[(q^{-1} - q)(t-2) - 1\bigr]^2\biggr)\biggr] \biggr)^{\alpha/2}.  \\
\end{split}
\end{equation}
In the stationary limit, one finds
\begin{equation}
\langle {\mathcal T}_\infty \rangle = \theta ~\frac{\Gamma(\frac{1+\alpha}{2})}{\sqrt{\pi}}  (2\gamma^2)^{\alpha/2} 
\biggl( \frac{2}{1+p} + \frac{2\beta_0^2 (1-q^2)}{(1-pq)(1+p)(1+q)}\biggr)^{\alpha/2} 
\end{equation}
that simplifies when $\lambda \ll 1$ and $\eta \ll 1$ to
\begin{equation}
\label{eq:TO}
\langle {\mathcal T}_\infty \rangle \approx \theta ~\frac{\Gamma(\frac{1+\alpha}{2}) 2^\alpha}{\sqrt{\pi}}  
 \biggl(\eta + \frac{\beta_0^2 \lambda \eta}{\lambda + \eta}\biggr)^{\alpha/2} 
\approx \theta ~\frac{\Gamma(\frac{1+\alpha}{2}) 2^\alpha}{\sqrt{\pi}} ~ \eta^{\alpha/2}  .
\end{equation}

\section{Variogram of incremental P\&L}
\label{sec:variogramPNL}

We sketch the derivation of the variogram of incremental P\&Ls in the
stationary limit $t_0\to\infty$.  The variogram is defined as
\begin{equation}
\label{eq:variogram_dPNL}
V_{t,t_0}^{\textrm{P\&L}} = \frac{\var\{ \dPNL_{t_0+1} + ... + \dPNL_{t_0+t}\}}{\var\{ \dPNL_{t_0+1}\} + ... + \var\{\dPNL_{t_0+t}\}} .
\end{equation}
The variances $\var\{\dPNL_k\}$ in the denominator are given by
Eq. (\ref{eq:vt}).  One can explicitly compute their sum for $k$
ranging from $t_0+1$ to $t_0+t$ and then take the limit
$t_0\to\infty$.  As expected, this limit is simply equal $t v_\infty$,
where the stationary variance $v_\infty$ is given by
Eq. (\ref{eq:v_infty}).  The major difficulties rely in the
computation of the numerator of Eq. (\ref{eq:variogram_dPNL}) which
contains correlations between incremental P\&Ls.

We start by writing the definition of the variance
\begin{equation}
\label{eq:auxil2}
\begin{split}
\var\{ \dPNL_{t_0+1} + ... + \dPNL_{t_0+t}\} & = \sum\limits_{j,k = t_0+1}^{t_0+t} 
\bigl[\langle r_j s_j r_k s_k\rangle - \langle r_j s_j\rangle \langle r_k s_k\rangle\bigr] \\
&= \sum\limits_{j,k = t_0+1}^{t_0+t} 
\bigl[\langle r_j r_k \rangle \langle s_j s_k\rangle + \langle r_j s_k\rangle \langle r_k s_j\rangle \bigr], \\
\end{split}
\end{equation}
where the second relation implied by the Wick's theorem.  Lengthy but
straightforward computation yields
\begin{equation*}
\begin{split}
\langle s_j s_k\rangle & = p^{|j-k|} - p^{j+k-2} +
\beta_0^2 \biggl(\frac{p(1-q^2)}{(1-pq)(p-q)} (p^{|j-k|} - p^{j+k-2}) \\
& - \frac{q(1-p^2)}{(1-pq)(p-q)} \bigl[p^{j+k-2} + q^{|j-k|} - p^{j-1} q^{k-1} - p^{k-1} q^{j-1}\bigr] \\
& - \frac{1-p^2}{(p-q)^2} (p^{j-1}-q^{j-1})(p^{k-1}-q^{k-1}) \biggr) \\
\end{split}
\end{equation*}
and
\begin{equation*}
\frac{\langle r_j s_k \rangle}{\gamma\beta_0^2} = \begin{cases} 
\frac{p^{k-j-1}}{\beta_0^2} + p^{k-j-1} \frac{1-(pq)^j}{1-pq} + 
q \frac{p^{k-j-1}-q^{k-j-1}}{p-q} - q^{j-1} \frac{p^{k-1}-q^{k-1}}{p-q} \quad (k > j), \cr
q^{j-k+1} \frac{1-(pq)^{k-1}}{1-pq} - q^{j-1} \frac{p^{k-1}-q^{k-1}}{p-q}  \hskip 46mm (k \leq j). \end{cases} 
\end{equation*}
The numerator of Eq. (\ref{eq:variogram_dPNL}) is then obtained by
computing the double sum in Eq. (\ref{eq:auxil2}).  Since we are
interested in the stationary limit $t_0\to\infty$, it is sufficient to
keep only the terms with $j-k$, in which the dependence on $t_0$ is
canceled.  In the stationary limit, one gets the following variance
\begin{equation}
\label{eq:var_PNL}
\begin{split}
& \lim\limits_{t_0\to\infty} \var\{\dPNL_{t_0+1} + ... + \dPNL_{t_0+t}\} = \biggl[1 + 2\beta_0^2 \frac{1+q^2-2p^2q^2}{(1-pq)^2} + \beta_0^4 c_2\biggr] t \\
& - 2\beta_0^2 q \frac{p+q-2p^2q}{(1-pq)^3} \biggl[1 + \beta_0^2 \frac{p(1-q^2)}{(1-pq)(p-q)} \biggr] \bigl(1 - (pq)^t\bigr) \\
& + \frac{4\beta_0^4 q^3(1-p^2)}{(1-pq)(p-q)(1-q^2)^2} \bigl(1 - q^{2t}\bigr),  \\
\end{split}
\end{equation}
where
\begin{equation*}
c_2 = \frac{2p^3q^5 + 2p^3q^3 - 6p^2q^2 + 2q^4p^2 - 4q^3p - q^5p + pq + 1 - q^4 + 4q^2}{(1-q^2)(1-pq)^3} .
\end{equation*}
For the special case $p=q$, one gets
\begin{equation*}
\begin{split}
& \lim\limits_{t_0\to\infty} \var\{\dPNL_{t_0+1} + ... + \dPNL_{t_0+t}\} = \biggl[1 + 2\beta_0^2 \frac{1+2q^2}{1-q^2} 
+ \beta_0^4 \frac{1+7q^2+2q^4}{(1-q^2)^2}\biggr] t \\
& - \frac{4\beta_0^2 q^2}{(1-q^2)^2}\biggl[1 + \beta_0^2 \frac{2q^2+3/2}{1-q^2}\biggr] (1 - q^{2t}) + \frac{4\beta_0^4 q^2}{(1-q^2)^2}~ q^{2t}~ t .  \\
\end{split}
\end{equation*}
Dividing Eq. (\ref{eq:var_PNL}) by $t v_\infty$, one gets the
variogram $V_{t,\infty}^{\textrm{P\&L}}$.  This expression can be
compared to the variogram of returns from Eq. (\ref{eq:Vst_exo}).
Both variograms behave similarly, exhibiting both rapid exponential
decay and slow power law decay.  The asymptotic value of the variogram
$V_{t,\infty}^{\textrm{P\&L}}$ as $t\to\infty$ is
\begin{equation}
V_{\infty,\infty}^{\textrm{P\&L}} = \frac{1 + 2\beta_0^2 \frac{1+q^2-2p^2q^2}{(1-pq)^2} + \beta_0^4 c_2}
{1 + \frac{2\beta_0^2}{1-pq} + \frac{\beta_0^4(1+q^2-2p^2q^2)}{(1-pq)^2}} .
\end{equation}
In the special case $p=q=1-\lambda$, one gets for $\lambda \ll 1$ and
$\beta_0 \ll 1$
\begin{equation}
V_{\infty,\infty}^{\textrm{P\&L}} \approx 1 + \frac{2\beta_0^2}{\lambda} + \frac{(\beta_0^2/\lambda)^2}{2(1 + \beta_0^2/\lambda)} .
\end{equation}
For comparison, the variogram of returns from Eq. (\ref{eq:Vst_exo})
gets the asymptotic value $1 + 2\beta_0^2/\lambda$.  It is worth
noting that $\beta_0^2/\lambda$ does not need to be a small parameter.

\section{The largest and the smallest eigenvalues at long times}
\label{sec:spectrum_long}

We present explicit formulas for the eigenvalues of the matrix $\M\C$
determining the cumulative P\&L for independent returns at long times.
In that case, $\M\C = \M_{1-\eta}^{(t,t_0)}$.  We first consider the
simpler case when an initiation period is not ignored (i.e., $t_0 =
0$).  In the limit $t\to\infty$, the matrix $\M_{1-\eta}^{(t,0)} =
\gamma[\A_{1-\eta} + \A_{1-\eta}^\T]$ becomes close to a cyclic matrix
whose eigenvalues can be computed as
\begin{equation}
\label{eq:spectrum_theory}
\mu_{1-\omega} = \gamma \sum\limits_{j=1}^\infty |p|^{j-1}\bigl[e^{i\pi j\omega} + e^{i\pi j\omega}\bigr] 
=  2\gamma \frac{\cos(\pi\omega) - p}{1 - 2p\cos(\pi\omega) + p^2}  ,
\end{equation}
where $\omega\in [0,1]$ is the ``index'', and $p = 1-\eta$.  The
largest eigenvalue of the limiting matrix is then
\begin{equation}
\label{eq:eigen_mup}
\mu_+^\infty = \mu_{\omega = 1} = \frac{2\gamma}{\eta} , 
\end{equation}
while the smallest eigenvalue is $\mu_{\omega = 0} = -
\frac{\gamma}{1-\eta/2}$.

When $t_0 > 0$, discarding the first $t_0$ points corresponds to
setting the block of size $t_0\times t_0$ of the matrix
$\M_{1-\eta}^{(t,0)}$ to zero.  This modification introduces $t_0$
zero eigenvalues into the spectrum and also changes the smallest
eigenvalue $\mu_-$ which becomes significantly smaller than
$\mu_{\omega=0}$.  In what follows, we compute the smallest eigenvalue
$\mu_-$ in the double limit $t_0\to\infty$ and $t\to\infty$.  For this
purpose, we first guess the corresponding eigenvector $\U$ and then
justify explicitly the correctness of the guess.  We take the vector
$\U$ of the form:
\begin{equation*}
\U = \bigl(\underbrace{p^{t_0-1}, p^{t_0-2}, ..., p, 1}_{t_0~\textrm{elements}}, -a, -a\p, -a\p^2, ...\bigr)^\T ,
\end{equation*} 
where $a$ and $\p$ are two parameters to be determined.  Applying the
matrix $\M_{1-\eta}^{(t,t_0)}$ to this vector, one gets
\begin{equation*}
\M_{1-\eta}^{(t,t_0)} \U = \gamma \left(\begin{array}{c} 
-a p^{t_0-1} (1 + p\p + ... ) \\
-a p^{t_0-2} (1 + p\p + ... ) \\
... \\
-a (1 + p\p + ... ) \\  \hline
(1 + p^2 + ... ) - a(\p[1 + p\p + ... ]) \\
p(1 + p^2 + ... ) - a(1 + \p^2[1 + p\p + ... ]) \\
...  \\  
p^{k+1}(1 + ... ) - a([p^k + p^{k-1}\p + ... + \p^k] + \p^{k+2}[1 + ... ]) \\
...  \\
\end{array} \right) ,
\end{equation*}
where the geometrical series $1 + p\p + (p\p)^2 + ...$ and $1 + p^2 +
p^4 + ...$ contain infinitely many terms in the limits $t\to\infty$
and $t_0\to\infty$, respectively.  If $\U$ is an eigenvector, the
right hand side has to be identified with $\mu_-^\infty \U$.  The
first $t_0$ identities read $\mu_-^\infty = \frac{-a\gamma}{1-p\p}$.
The identity in the $(t_0+1)$-th row gives $-a\mu_-^\infty =
\gamma[\frac{1}{1-p^2} - \frac{a\p}{1-p\p}]$, from which
\begin{equation}
\label{eq:auxil1}
\frac{a(a+\p)}{1-p\p} = \frac{1}{1-p^2} .
\end{equation}
Finally, an identity in the $(t_0 + k+2)$-th row reads
\begin{equation*}
-a \p^{k+1} \mu_-^\infty = \gamma\biggl[\frac{p^{k+1}}{1-p^2} - a(p^k + p^{k-1}\p + ... + \p^k) - \frac{a\p^{k+2}}{1-p\p}\biggr] ,
\end{equation*}
from which one expresses $\p = p - a(1-p^2)$.  Substituting this
relation into Eq. (\ref{eq:auxil1}), one gets $a = 1/p$, from which
$\p = 2p - 1/p$ and finally
\begin{equation}
\label{eq:eigen_mum}
\mu_-^\infty = - \frac{\gamma}{2p(1-p^2)}.
\end{equation}
In this way, we constructed explicitly an eigenvalue $\mu_-^\infty$
and the corresponding eigenvector $\U$ of the limiting matrix
$\M_{1-\eta}^{(\infty,\infty)}$.  In turn, we did not show that
$\mu_-^\infty$ is the smallest eigenvalue.  Although the related
demonstration could in principle be performed, this analysis is beyond
the scope of the paper.  We checked numerically that $\mu_-^\infty$
from Eq. (\ref{eq:eigen_mum}) accurately approximates the smallest
eigenvalue of the matrix $\M_{1-\eta}^{(t,t_0)}$ for large enough
$t_0$ and $t$.


\begin{thebibliography}{35}

\bibitem{Covel}          M. W. Covel,
                         Trend Following (Updated Edition): Learn to Make Millions in Up or Down Markets,
                         Pearson Education, New Jersey, 2009.                 

\bibitem{Clenow}         A. F. Clenow,
                         Following the Trend: Diversified Managed Futures Trading,
                         Wiley \& Sons, Chichester UK, 2013.

\bibitem{Chan96}         L. K. C. Chan, N. Jegadeesh, J. Lakonishok,
                         J. Finance {\bf 51} (1996) 1681. 

\bibitem{Jegadeesh01}    N. Jegadeesh, S. Titman,
                         J. Finance {\bf 56} (2001) 699. 

\bibitem{Chan99}         L. K. C. Chan, N. Jegadeesh, J. Lakonishok,
                         Finan. Anal. J. {\bf 55} (1999) 80. 


\bibitem{Moskowitz12}    T. J. Moskowitz, Y. H. Ooi, L. H. Pedersen,
                         J. Finan. Econ. {\bf 104} (2012) 228. 



\bibitem{Asness13}       C. S. Asness, T. J. Moskowitz, L. H. Pedersen,
                         J. Finance {\bf 68} (2013) 929. 




\bibitem{Vandewalle98}   N. Vandewalle, M. Ausloos,
                         Phys. Rev. E {\bf 58} (1998) 6832. 

\bibitem{Vandewalle99}   N. Vandewalle, A. Ausloos, P. Boveroux,
                         Physica A {\bf 269} (1999) 170.

\bibitem{Carbone04}      A. Carbone, G. Castelli, H. E. Stanley,
                         Phys. Rev. E {\bf 69} (2004) 026105.

\bibitem{Arianos11}      S. Arianos, A. Carbone, C. T\"urk,
                         Phys. Rev. E {\bf 84} (2011) 046113.





\bibitem{Potters05}      M. Potters, J.-P. Bouchaud,
                         Wilmott Magazine (Jan 2006); online: ArXiv physics-0508104 (2005).

\bibitem{Martin12}       R. Martin, D. Zou,
                         Momentum trading: 'skews me, 
                         Risk Magazine (2012).


\bibitem{Andersen00}     T. G. Andersen, T. Bollerslev, F. X. Diebold, P. Labys,
                         Multinat. Finance J. {\bf 4} (2000) 159. 



\bibitem{Bouchaud}       J.-P. Bouchaud, M. Potters,
                         Theory of Financial Risk and Derivative Pricing: From Statistical Physics to Risk Management,
                         Cambridge University Press, 2003.

\bibitem{Mantegna}       R. Mantegna, H. E. Stanley, 
                         An introduction to Econophysics,
                         Cambridge University Press, Cambridge, 1999.

\bibitem{Mantegna95}     R. Mantegna, H. E. Stanley,
                         Nature {\bf 376} (1995) 46. 

\bibitem{Gabaix03}       Gabaix, Gopikrishnan, Plerou, Stanley,
                         Nature {\bf 423} (2003) 267. 

\bibitem{Bouchaud01}     J.-P. Bouchaud, M. Potters,
                         Physica A {\bf 299} (2001) 60.

\bibitem{Sornette03}     D. Sornette,
                         Phys. Rep.  {\bf 378} (2003), 1. 

\bibitem{Bouchaud04}     J.-P. Bouchaud, Y. Gefen, M. Potters, M. Wyart, 
                         Quant. Finance {\bf 4} (2004) 176. 

\bibitem{Stella08}       A. L. Stella, F. Baldovin,
                         Pramana J. Phys. {\bf 71} (2008) 341.

\bibitem{Bouchaud01b}    J.-P. Bouchaud, A. Matacz, M. Potters, 
                         Phys. Rev. Lett. {\bf 87} (2001) 1. 

\bibitem{Valeyre13}      S. Valeyre, D. S. Grebenkov, S. Aboura, Q. Liu, 
                         Quant. Finance (in press).







\bibitem{Box}            G. Box, G. M. Jenkins, G. C. Reinsel, 
                         Time Series Analysis: Forecasting and Control, 
                         Third ed., Prentice-Hall, 1994.


\bibitem{Holt57}         C. C. Holt, 
                         Office of Naval Research Memorandum 52 (1957); reprinted in 
                         Int. J. Forecast. {\bf 20} (2004) 5. 

\bibitem{Winters60}      P. R. Winters, 
                         Management Science {\bf 6} (1960) 324. 

\bibitem{Brown}          R. G. Brown, 
                         Smoothing Forecasting and Prediction of Discrete Time Series,
                         Englewood Cliffs, NJ: Prentice-Hall, 1963.





\bibitem{Grebenkov11}    D. S. Grebenkov,  
			 Phys. Rev. E {\bf 84} (2011) 031124.

\bibitem{SM}             Supplementary Materials.








\end{thebibliography}
\end{document}